\documentclass{ar2e}
\usepackage{graphicx}
\usepackage{latexsym}
\usepackage{morefloats}
\usepackage{amssymb,amsmath}

\input epsf.tex    

\newcommand{\Dz}{D^0}
\newcommand{\Dzb}{\overline{D}^0}
\newcommand{\Dzbar}{\overline{D}^0}
\newcommand{\barD}{\overline{D}^0}
\newcommand{\DzDzb}{D^0-\overline{D}^0}

\newcommand{\Kp}{K^+}
\newcommand{\Km}{K^-}
\newcommand{\KS}{K_S}
\newcommand{\KL}{K_L}
\newcommand{\pim}{\pi^-}
\newcommand{\pip}{\pi^+}
\newcommand{\piz}{\pi^0}
\newcommand{\etap}{\eta^\prime}

\newcommand{\kappares}{\kappa}
\newcommand{\sigmares}{\sigma}

\newcommand{\bea}{\begin{eqnarray}}
\newcommand{\eea}{\end{eqnarray}}
\newcommand{\beq}{\begin{equation}}
\newcommand{\eeq}{\end{equation}}

\newcommand{\CP}{\ensuremath{C\!P}}
\newcommand{\CPV}{\ensuremath{C\!PV}}
\newcommand{\babar}{B{\small A}B{\small AR}}
\newcommand{\ifb}{fb$^{-1}$}

\newcommand{\mco}{\multicolumn}
\newcommand{\clr}{\phantom{.}}

\newcommand{\mup}{\ensuremath{\mu^+}}
\newcommand{\mum}{\ensuremath{\mu^-}}

\newcommand{\ep}{\ensuremath{e^+}}

\newcommand{\Dp}{\ensuremath{D^+}}
\newcommand{\Dm}{\ensuremath{D^-}}
\newcommand{\Dsp}{\ensuremath{D_s^+}}
\newcommand{\Ds}{\ensuremath{D_s}}
\newcommand{\Dsstar}{\ensuremath{D_s^\star}}
\newcommand{\fd}{$ f_{D^+}$}
\newcommand{\fds}{$f_{D_s^+}$}
\newcommand{\rfd}{$f_{D_s^+}/f_{D^+}$}
\newcommand{\etal}{{\em et\ al.}}

\newcommand{\epem}{e^+e^-}
\newcommand{\epm}{e^+e^-}

\newcommand{\ampl}{A}
\newcommand{\asymd}{\ensuremath{a_{\hbox{\tiny D}}}}
\newcommand{\asymm}{\ensuremath{a_{\hbox{\tiny M}}}}

\newcommand{\RD}{\ensuremath{R_{\hbox{\tiny D}}}}
\newcommand{\RM}{\ensuremath{R_{\hbox{\tiny M}}}}
\newcommand{\xd}{\ensuremath{x_{\hbox{\tiny D}}}}
\newcommand{\yd}{\ensuremath{y_{\hbox{\tiny D}}}}
\newcommand{\ycp}{\ensuremath{y_{\hbox{\tiny D}}^{\scriptscriptstyle CP}}}
\newcommand{\gamd}{\ensuremath{\Gamma_{\hbox{\tiny D}}}}
\newcommand{\emd}{\ensuremath{M_{\hbox{\tiny D}}}}
\newcommand{\rCS}{\ensuremath{r_{\hbox{\tiny CS}}}}
\newcommand{\rDCS}{\ensuremath{r_{\hbox{\tiny DCS}}}}
\newcommand{\ie}{{\sl i.e.}}

\begin{document}

\title{Charm Meson Decays}

\markboth{Artuso, Meadows, and Petrov}{Charm Meson Decays}

\author{Marina Artuso$^a$, Brian Meadows$^b$, Alexey A Petrov$^{c,d}$
\affiliation{$^a$Syracuse University, Syracuse, NY 13244, USA\\
$^b$University of Cincinnati, Cincinnati, OH 45221, USA \\
$^c$Wayne State University, Detroit, MI 48201, USA \\
$^d$MCTP, University of Michigan, Ann Arbor, MI 48109, USA
}}

\begin{keywords}
charmed quark, charmed meson, weak decays, CP-violation
\end{keywords}

\begin{abstract}
We review some recent developments in charm meson physics. In
particular, we discuss theoretical predictions and experimental
measurements of  charmed meson decays to leptonic, semileptonic, and
hadronic final states and implications of such measurements to
searches for new physics. We discuss $\DzDzb$-mixing and
CP-violation in charm, and discuss future experimental prospects and
theoretical challenges in this area.
\end{abstract}

\maketitle

\section{INTRODUCTION}\label{Intro}

The charm quark has played a unique role in particle physics for
more than three decades.  Its discovery by itself was an important
validation of the Standard Model (SM), as its existence and mass
scale were predicted~\cite{Gaillard:1974mw} on the basis of
low-energy kaon experiments before any direct experimental signature
for charm was available.

Several features distinguish charmed hadrons from those with other flavors.
While their mass,
${\cal O}(2\mbox{~GeV})$ places them in the region where
non-perturbative hadronic physics is operative, theoretical methods
developed for heavy quarks can in principle still be applied, albeit
with larger uncertainties. On the other hand, recent advances in
unquenched lattice QCD simulations paved the way for charm data to
be used to probe the Yukawa sector of the SM. It is the
only up-type quark that can have flavor oscillations. Finally, charm
decays provide a unique window on new physics (NP) affecting the up-type
quark dynamics. In many cases, charm transitions provide almost
background-free low-energy signals of NP. For example, many
popular NP models predict signals for CP-violation (\CPV)
much larger than what is generally predicted within the
SM~\cite{petrov}.
One hopes that, just like manifestation for charm quark existence
came from low-energy kaon oscillation experiments, oscillations of
charmed hadrons can provide hints of what is happening at the TeV
scale.

Experiments operating at the $\psi(3770)$ resonance, near threshold
for $D\bar{D}$ production, such as MARK III at SPEAR, performed the
initial exploration of charm phenomenology \cite{markiii}. Later,
higher energy machines, either fixed target experiments operating at
hadron machines or higher energy $\epem$ colliders, entered this
arena, with much bigger data samples. In recent years, we have seen
a renewed interest in studying open charm in $\epem$ colliders with
a center-of-mass energy close to $D\bar{D}$ threshold. The CLEO
experiment\cite{yellowbook} at CESR, after years of charm studies at
a center-of-mass energy near the $\Upsilon (4S)$, has collected a
sample exceeding 800 pb$^{-1}$ at the $\psi(3770)$ center-of-mass
energy, and a sample of about 600 pb$^{-1}$ at a center-of-mass
energy close to 4170 MeV, optimal for $\Ds$ studies. The BES-II
experiment, at BEPC, has published results based on 33 pb$^{-1}$
accumulated around the $\psi(3770)$. It has an upgrade program both
for the detector (BESIII) and the machine (BEPCII), designed as a
charm factory with $10^{33} {\rm cm^{-2}s^{-1}}$ peak luminosity
that should be completed in 2008 \cite{bes3}. In parallel, \babar\
and Belle have provided important contributions to our knowledge of
charm decays, exploiting their impressive data sets. Both $B$
factory facilities have achieved luminosities almost ten times their
original design. A KEK-B upgrade has been approved to attain
luminosities about one order of magnitude greater than already
achieved. An alternative approach, with higher luminosity goals and
the added capability to run at both $\Upsilon(4S)$ and at the
$\psi(3770)$, is being considered, but is not approved. Finally, CDF
and D0 have entered the arena of charm physics, applying to this
study some of the tools developed to pursue beauty physics.

Experiments operating at the $\psi(3770)$ resonance have several
advantages: the final state is extremely simple, being dominated by
a $D\bar{D}$ pair. The cross section for charm production is
relatively high, $\sigma (D^0\bar{D}^0)=3.66 \pm 0.03\pm 0.06$~nb
and $\sigma (D^+D^-)=2.91 \pm 0.03\pm 0.05$~nb~\cite{:2007zt}. In
addition, the relatively high branching fractions to low
multiplicity final states allow the use of tagged samples, where one
$D$ is fully reconstructed (tag) and the rest of the event recoiling
against the tag is used to study specific signals. This technique is
particularly useful to study leptonic and semileptonic events, as it
allows a precise reconstruction of the neutrino 4-momentum, and of
the invariant mass squared of the lepton-neutrino pair ($q^2$). In
addition, the $D\bar{D}$ pairs are produced in a $C=-1$ state, and
this quantum coherence allows unique probes of mixing and $CPV$. On
the other hand the b-factory experiments, Belle and BaBar, can also
be considered charm factories. At the $\Upsilon\mbox{(4S)}$
center-of-mass energy, $\sigma(b\bar{b})\sim 1.1$ nb, while
$\sigma(c\bar{c})\sim 1.3$ nb.  The background to be tackled is
higher than at the lower energy, and neutrino and $q^2$
reconstruction in general are not as precise. Significant
improvements are achieved, however, through the use of tagged
samples \cite{Widhalm:2006wz}, made possible by the impressive size
of the data set accumulated: in January 2008 the samples are 484
fb$^{-1}$ at \babar\ and 763 fb$^{-1}$ at Belle. Also, the energy at
which they operate makes possible the production of charmed baryons
and boosts the charm particles sufficiently so that time dependent
measurements are possible.

Experiments at hadron machines have the advantage of much higher
cross sections for charm and beauty production, at the expense of
significant backgrounds. Luckily the relative long lifetime of charm
hadrons ($\sim 1$ ps), combined with the development of silicon
micropattern detectors provides a unique and powerful signature of
charm meson decays: the identification of detached secondary
vertices.  About 30 years after its initial inception \cite{na1},
this technique is still being perfected, introducing vertex
algorithms more and more sophisticated that provide charm and beauty
event tagging almost in real time. This development has allowed
fixed target experiments and the two Tevatron experiments, CDF and
D0, to provide significant contributions to our knowledge of charm
decays, with competitive limits on some rare decays and recent
results in $\Dz\Dzb$ mixing. This work will be continued by LHCb,
the first dedicated charm and beauty experiment at a hadron collider
(LHC), that relies heavily on detached vertex criteria in early
stages of its triggering process and is considering an upgrade that
would include detached vertex criteria in the lowest trigger level
\cite{ma-vertex}.

\section{LEPTONIC AND SEMILEPTONIC DECAYS} \label{Lept}

Charm leptonic and semileptonic decays are ideal laboratories to
study non-perturbative QCD, and to determine important quark mixing
parameters. In addition, they may provide additional constraints on
physics beyond the SM.

In the SM semileptonic decays are described by an effective
Hamiltonian
\beq\label{Hsl}
{\cal H}=\frac{G_F}{\sqrt{2}} V_{cq} L^\mu
\bar q\Gamma_\mu c,
\eeq
with $L^\mu$ being a leptonic current, $G_F$ is a Fermi constant, and
$\Gamma_\mu=\gamma_\mu \left(1-\gamma_5\right)$. Theoretically,
leptonic decays are the simplest to describe, as they only depend
on a single non-perturbative parameter, the decay constant $f_{D_q}$,
\beq
\langle 0 | \bar q\gamma^\mu \gamma_5 c | D_q \rangle = -i f_{D_q} p_D^\mu,
\eeq
which parameterizes the amplitude of probability for the heavy and a
light quark to ``find each other'' in a meson. Semileptonic decays
are traditionally  described in terms of form factors cast as a
function of $q^2$, the invariant mass of the electron-neutrino pair.
Experimental determinations of these form factors are performed
through the study of the differential decay width $d\Gamma /dq^2$.
In both cases, decay constants and form factors are  QCD parameters
that can only be computed using non-perturbative techniques.

Accurate calculations of non-perturbative QCD parameters are very
challenging. Lattice QCD  represents an appealing approach. In
principle, it is the only one that can be improved in a systematic
way. A big stumbling block has been the inclusion of dynamical quark
effects (unquenched lattice QCD). Recently, technical developments
such as highly improved actions of QCD and the availability of ``2+1
flavor'' MILC configurations with 3 flavors of improved staggered
quarks have lead to results with much higher accuracy and allowed
for consistent estimate of both statistical and systematic errors
involved in the simulations. Two groups have reported  charm decay
constant calculations with three dynamical quark flavors: the
Fermilab/MILC Lattice collaboration~\cite{Aubin:2005ar} and the
HPQCD collaboration~\cite{Follana:2007uv}. They both use the ``2+1''
MILC configurations including three flavors of staggered quarks: one
heavier with a mass close to the strange quark mass $m_s$, and two
degenerated light quarks with masses  between $m_s/10$ and $m_s$,
but differ on how they treat heavy quarks in their formulations of
lattice QCD. The Fermilab group has also calculated the shape and
normalization of the form factors in semileptonic $D \to \pi e \nu
_e$ and $D\to K e \nu _e$ decays \cite{Aubin:2004ej}.

The QCD sum rules \cite{Shifman:1978bx,Narison:2002hk} provide a
method for calculating hadronic matrix elements including
non-perturbative effects that was designed to make maximum use of
the known manifestations of non-perturbative QCD. A few parameters
describe the non-pertubative dynamics and are fixed from well known
hadronic processes, and then applied to heavy meson decays. Finally,
quark models, generally QCD inspired and based on a variety of
assumptions, have been used to predict form factor normalizations
and decay constants~\cite{Scora:1995ty}.

In principle, charm meson semileptonic decays provide the simplest
way to determine the magnitude of quark mixing parameters: the charm
sector allows direct access to $|V_{cs}|$ and $|V_{cd}|$.
Semileptonic decay rates are related to $|V_{cq}|^2$ via matrix
elements that describe strong interaction effects.

The study of charm semileptonic decays may contribute to a precise
determination of the  Cabibbo Kobayashi Maskawa ($CKM$) matrix
element $|V_{ub}|$. A variety of theoretical approaches have been
proposed to use constraints provided by charm decays to reduce the
model dependence in the extraction of $|V_{ub}|$ from exclusive
charmless $B$ semileptonic decays. In particular, if heavy quark
effective theory (HQET) \cite{hqet} is applicable both to the $c$
and $b$ quarks, there is an SU(2) flavor symmetry that relates the
form factors in $D$ and $B$ semileptonic decays \cite{isgur-wise}.
For example, a flavor symmetry relates the form factors in
$D\rightarrow \pi \ell \nu$ are related to the ones in $B\rightarrow
\pi \ell \bar{\nu}$, at the same $E\equiv{v}\cdot {p}_{\pi}$,  where
$E$ is the energy of the light meson in the center-of-mass $D$ frame,
${v}$ is the four-velocity of the $D$ meson, and ${p}_P$ is the 4-momentum of
the light hadron. The original method has been
further refined~\cite{ben}; the large statistics needed to implement
these methods may be available in the near future.

\subsection{Theoretical Predictions for the Decay Constant}

The leptonic decay width is given by
\begin{equation}\label{fd:eq}
\Gamma(D_q\to \ell\nu) = {G_F^2\over
8\pi}f_{D_q}^2m_{\ell}^2M_{D_q}  \left(1-{m_{\ell}^2\over
M_{D_q}^2}\right)^2 \left|V_{cq}\right|^2~~~,
\end{equation}
where $q=d,s$ for $D^+$ or $D_s$ states respectively, $M_{D_q}$ is
the $D_q$ mass, $m_{\ell}$ is the mass of the final state lepton,
and $|V_{cq}|$ is the $CKM$ matrix element associated with the $c
\to q$ transition. Due to helicity suppression, the rate goes as
$m_\ell^2$; consequently the electron mode $D^+ \to e^+\nu _e$ has a
very small rate in the SM. The relative widths scale as
$2.65:1:2.3\times 10^{-5}$ for the $\tau^+ \nu_{\tau}$, $\mu^+
\nu_{\mu}$ and $e^+ \nu _e$ final states, respectively. The decay
constant $f_D$ parameterizes a matrix element of the axial current
and is the only non-perturbative parameter in Eq.~(\ref{fd:eq}). It
can be related to the wave function overlap of charm quark and light
antiquark. Charm meson decay constants are amenable to experimental
and theory determination both for $\Dp$ and $\Ds$, thus allowing a
direct measurement of SU(3) breaking and a comparison with the
theory.
\begin{table}
\caption{Theoretical predictions for \fd ,\fds ,\rfd
.\label{tab:fdsum}}
\begin{tabular}{lccc}
\hline Authors & \fd\ (MeV)& \fds (MeV) & \rfd\\\hline
\multicolumn{4}{c}{Unquenched lattice calculations}\\ \hline
HPQCD+UKQCD \cite{Follana:2007uv} & $208\pm 4$ & $241\pm 3$ &$1.162 \pm 0.009$\\
FNAL+MILC+HPQCD \cite{Aubin:2005ar} & $201\pm 3 \pm 17$ & $249 \pm 3
\pm 16$ & $1.24\pm 0.01 \pm 0.07$ \\\hline
\multicolumn{4}{c}{Quenched Lattice QCD Calculations}\\\hline Taiwan
\cite{chiu}& $235\pm 8\pm
14 $ & $266\pm 10\pm 18$ & $1.13 \pm 0.03 \pm 0.05$ \\
UKQCD \cite{ql:lellouch} & $210 \pm 10 ^{+17}_{-16}$ & $236 \pm 8
^{+17}_{-14}$ & $1.13
\pm 0.02 ^{+0.04}_{-0.02}$\\
Becirevic \etal\cite{ql:becirevic} & $211\pm 14 ^{+2}_{-12}$ & $231
\pm 12 ^{+6}_{-1}$ & $1.10\pm 0.02$\\\hline \multicolumn{4}{c}{QCD
sum rules and other approximations}\\\hline J. Bordes \etal
\cite{fd:bordes}& $177\pm 21$ & $205\pm 22$
& $1.16\pm 0.02\pm 0.03$\\
S. Narison  \cite{fd:narison} & $203\pm 10$ & $235\pm 24$ & $1.15\pm 0.04$\\
Field Correlators \cite{badalia} & $210\pm 10$ & $260\pm 10$ & $1.24 \pm 0.03$\\
Isospin Splitting \cite{amundson} & ~~~ & $262\pm 29$ & ~~
\\\hline
\end{tabular}
\end{table}

\subsection{Experimental Determinations of $f_D$}

\begin{figure}
\epsfxsize 25pc
\centerline{\epsfbox{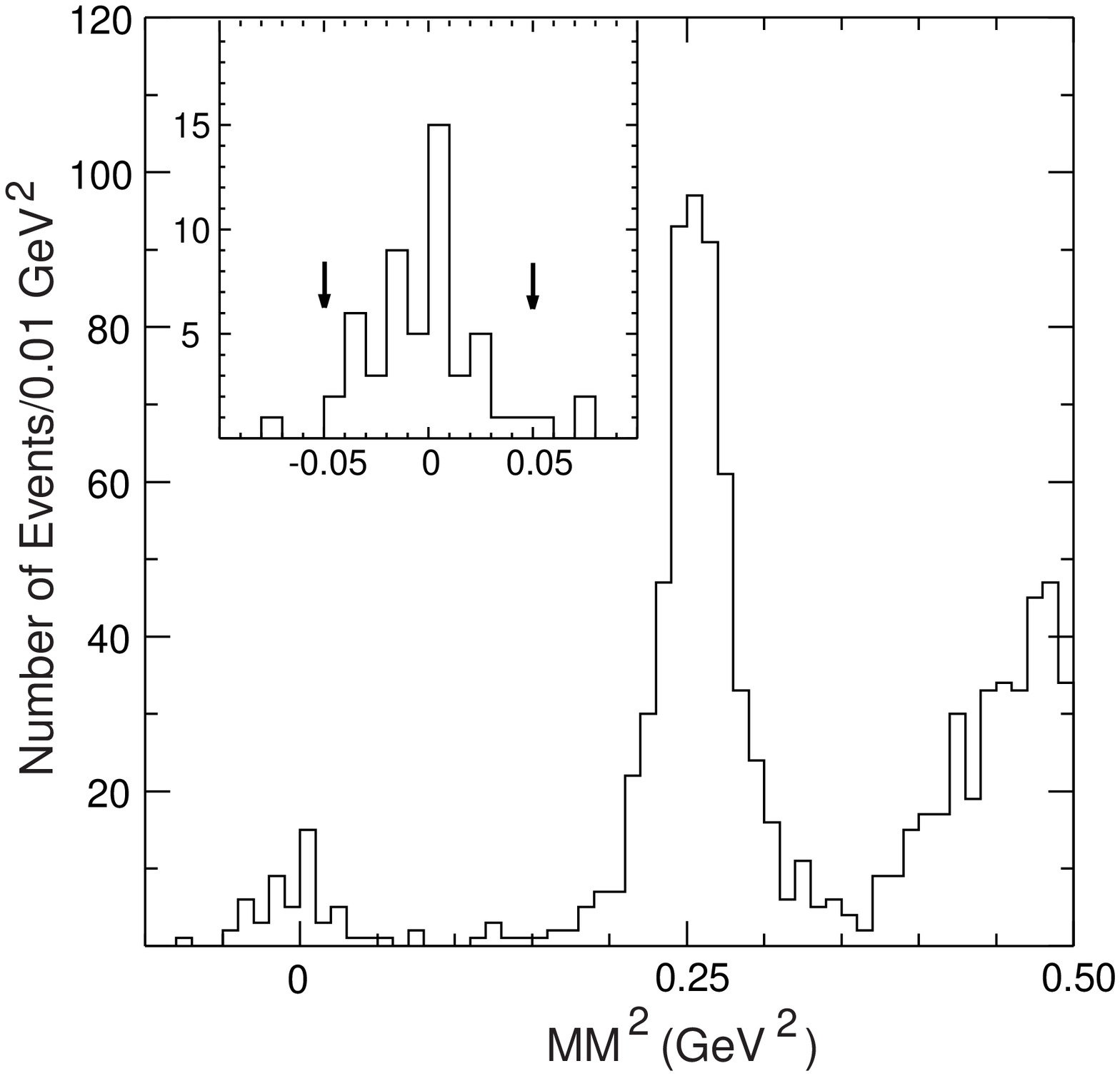}}
\caption{CLEO-c
$MM^2$ using $D^-$ tags and one opposite charged track with no extra
energetic clusters.$^{10}$ The insert shows the signal region for
$D^+\rightarrow \mu\nu_{\mu}$ enlarged; the defined signal region is
shown between the two arrows.} \label{fd:data}
\end{figure}

The CLEO collaboration \cite{Artuso:2005ym} has measured $f_{D^+}=
(222.6 \pm 16.7 ^{+2.8}_{-3.4})$ MeV, using a tagged sample of
$\Dp\Dm$ decays collected at a center-of-mass energy close to 3.77
GeV. The existence of the neutrino is inferred by requiring the
missing mass squared ($MM^2$) to be consistent with zero,
\begin{equation}
MM^2= (E_{beam}-E_{\mu^+})^2-(\vec{p}_{D^-}-\vec{p}_{\mu
^+})^2,\nonumber
\end{equation}
Figure \ref{fd:data} shows the measured $MM^2$, with a 50 event peak
in the interval [-0.050 GeV$^2$,+0.050 GeV$^2$], approximately $\pm
2\sigma$ wide. The background is evaluated as $2.81\pm 0.30\pm 0.27$
events. The same tag sample is used to search for $D^+\rightarrow
e^+ \nu_{e}$. No signal is found, corresponding to a 90\% CL upper
limit ${\cal B}(D^+\rightarrow e^+ \nu _e)< 2.4\times 10^{-5}$.
%
%
More data is available on \fds . Early measurements and a recent
BaBar result of \fds\ determine the ratio ${\cal B}(\Ds \to
\mu\nu)/{\cal B}(\Ds \to \phi \pi)$ \cite{Aubert:2006sd}. This adds
an additional large source of error as the denominator is not well
known~\cite{Yao:2006px}.  CLEO-c uses a sample of $\Ds\Dsstar$
collected near the center-of-mass energy of 4.17 GeV to study $\Ds$
leptonic decays \cite{Artuso:2007zg}. They measure the branching
fraction for the decays $\Ds\to \mu\nu_{\mu}$, $\Ds\to \tau
\nu_{\tau}$, with $\tau \to \pi \nu_{\tau}$, and $\tau \to
e\nu_{\tau}\nu_e$ \cite{Ecklund:2007zm}. Recently, Belle has
reported an absolute value for the branching fraction ${\cal B}(\Ds
\to \mu \nu_{\mu})$ based on fully reconstructed samples of events
of the type $\epm \to \Dsstar DKX$, $\Dsstar \to \Ds\gamma$ where
$X$ is any number of $\pi$ and at most one $\gamma$ from
fragmentation \cite{:2007ws}.  Theoretical predictions are
summarized in Table \ref{tab:fdsum}, while measurements are shown in
Table \ref{fds:tab}. The average of the absolute measurements is
\fds\ = 275 $\pm$ 10 MeV, assuming that $|V_{cs}|=|V_{cd}|=0.9737$.
Typically, the experimental value is above theoretical predictions.
In general, the errors are such that the discrepancy is not yet
meaningful, with the exception of the most recent UKQCD-MILC
calculation \cite{Follana:2007uv}. In this case, the discrepancy
between theory and experiment exceeds their stated errors by about 3
$\sigma$.
\begin{table}[hbtp]
\caption{Results for ${\cal B}_{\phi\pi}\equiv {\cal B}(\Ds\to
\mu^+\nu_{\mu})$, ${\cal B}(\Ds\to \tau^+\nu_{\tau})$, and
$f_{\Ds^+}$. (Numbers have been updated using $\Ds$ lifetime of 0.50
ps.) Results below the line have not been used in this average from
Ref.~\cite{stone:pdg}. The assumed value of ${\cal
B}(\Ds^+\to\phi\pi^+)$ is listed whenever available. ALEPH average
their two results to obtain a value for \fds . \label{fds:tab}}
\begin{tabular}{llccc}
\hline Exp. & Mode & ${\cal B}(x10^3)$ & ${\cal
B}_{\phi\pi}(\%)$ & \fds\ (MeV)\\\hline CLEO-c & $\mu ^+\nu_{\mu}$
\cite{Artuso:2007zg} &$5.94\pm 0.66\pm 0.31$ & ~~ & $264\pm 15 \pm
7$\\
CLEO-c & $\tau ^+\nu_{\tau}$ \cite{Artuso:2007zg} & $80.0\pm 13.0
\pm 4.0)$& ~~ &$310\pm 25\pm 8$\\ CLEO-c & $\tau ^+\nu_{\tau}$
\cite{Ecklund:2007zm} & $61.7 \pm 7.1 \pm 3.6 $ & ~~ &  $275 \pm 10
\pm 5$ \\
CLEO-c & combined & ~~ &~~ & $274 \pm 10 \pm 5$\\
Belle & $\mu ^+\nu_{\mu}$\cite{:2007ws} & $6.44\pm 0.76\pm 0.52 $ &~~& $279\pm16 \pm 12$\\
Average & ~~ & ~~ & ~~& $275\pm 10$ \\
\hline CLEO \cite{Chadha:1997zh}&$\mu ^+\nu_{\mu}$& $6.2\pm 0.8\pm
1.3 \pm 1.6$ &
$3.6\pm 0.9$ & $273 \pm 19\pm 27 \pm 33$ \\
BEATRICE \cite{Alexandrov:2000ns}& $\mu ^+\nu_{\mu}$ & $8.3\pm
2.3\pm 0.6\pm 2.1$& $3.6\pm 0.9$
& $312\pm 43\pm 12\pm 39$\\
ALEPH \cite{Heister:2002fp} & $\mu ^+\nu_{\mu}$ & $6.8\pm 1.1\pm 1.8
$& $3.6\pm 0.9$ & $ 282\pm 19 \pm 40$ \\
ALEPH \cite{Heister:2002fp} & $\tau ^+\nu_{\tau}$ & $58\pm 8 \pm 18$
& ~~ & ~~ \\
L3 \cite{Acciarri:1996bv} & $\tau ^+\nu_{\tau}$ & $74 \pm 28 \pm
16\pm 18$ & ~~ & $299\pm 57\pm 32 \pm 37$\\
OPAL \cite{Abbiendi:2001nb} & $\tau ^+\nu_{\tau}$ & $70\pm 21\pm 20$
& ~~ & $283\pm 44 \pm 41$ \\
BaBar \cite{Aubert:2006sd} & $\mu ^+\nu_{\mu}$ & $6.74\pm 0.83\pm
0.26\pm 0.66$ & $4.71\pm 0.46$ & $283\pm 17 \pm 7 \pm
14$\\\hline
\end{tabular}
\end{table}

\subsection{Constraints on New Physics from $f_D$}

Leptonic decays are sensitive probes of NP interactions mediated by
charged particles. Models with an extended Higgs sector, which
include new charged scalar states, or models with broken left-right
symmetry, which include massive vector $W^\pm_R$ states, are primary
examples of such interactions. Recent evidence of observation of
$B\to \tau\nu_{\tau}$ decay brought renewed attention to such
models. In particular, two Higgs doublet models, including Minimal
Supersymmetric SM (MSSM), could give contributions to such
transitions. Different implementations of this extension of the SM
can be formulated~\cite{hewett:newphys}. For example, the first
doublet ($\Phi_1$) could give mass to the up-type fermions and the
second ($\Phi_2$) to the down-type fermions. In this case,
\begin{eqnarray}
{\cal B}(\Dp\to \ell^+ \nu_{\ell}) &=&
{\cal B} _{SM}\left( 1+\frac{m_D^2}{m^2_{H^\pm}}\right)^2 \\
\nonumber {\cal B}(\Dsp\to \ell^+ \nu_{\ell}) &=&  {\cal B}_{SM}
\left[1+\frac{m_{\Ds}^2}{m^2_{H^\pm}}
\left(1-\tan^2{\beta}\frac{m_s}{m_c}\right)\right]^2
\end{eqnarray}
Note that the latter model introduces a correction to the SM
expectations that may be considerable and negative at large
$\tan^2{\beta}$. A limit can also be set on the mass of a charged Higgs,
$m_{H^+} > 2.2\tan{\beta}$. This limit is similar to the one obtained from the
measurement ${\cal B}(B\to \tau\nu)$ decay \cite{belle:fb}.

\subsection{Absolute Branching Fractions for Semileptonic $D$ Decays}

\begin{figure}
\epsfxsize 25pc
\centerline{\epsfbox{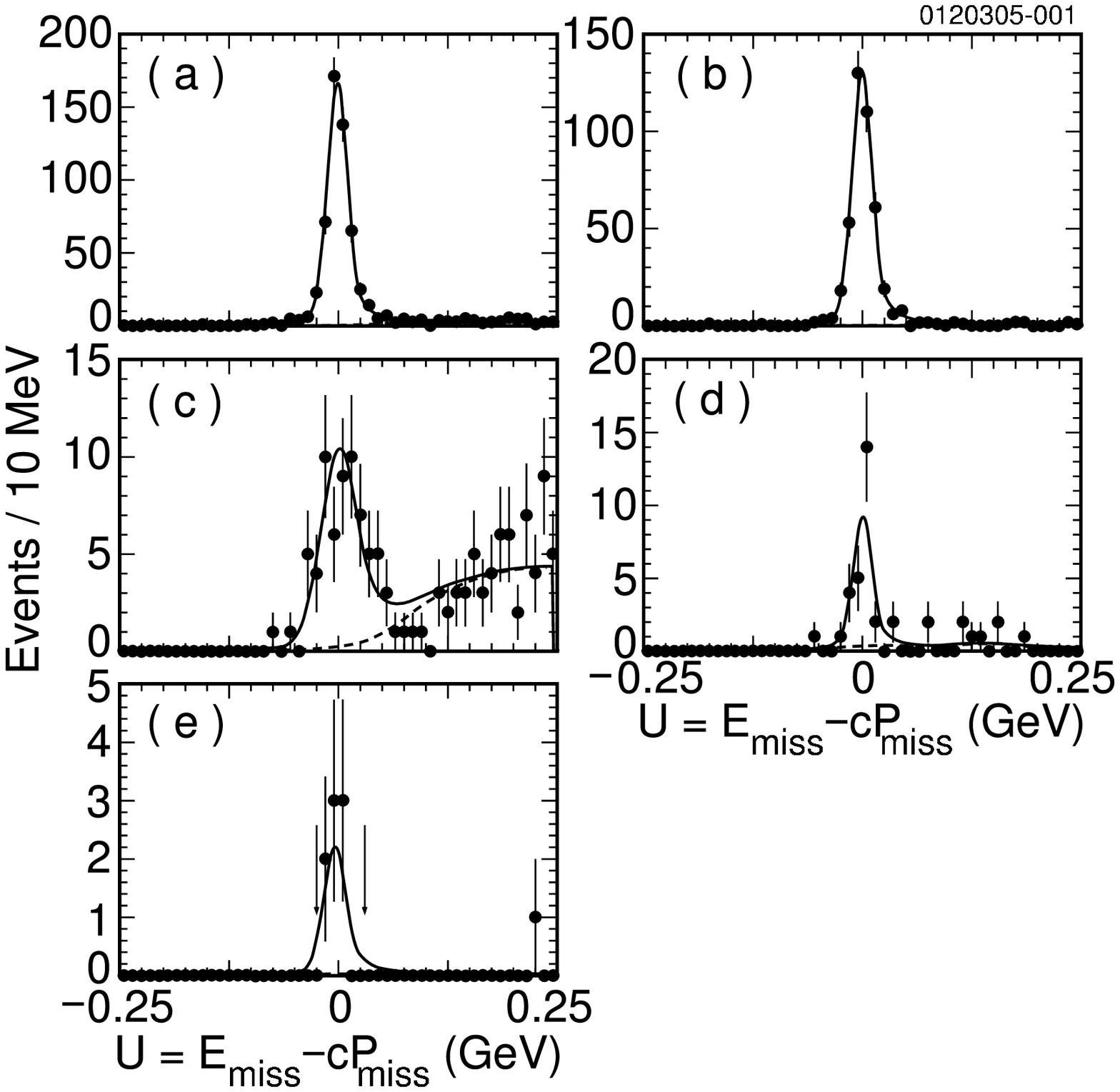}}\
\caption{Fits (solid
lines) to the $U$ distributions in CLEO-c$^{28}$ data (dots with
error bars) for the five $D^+$ semileptonic modes: (a)
$D^+\rightarrow \bar{K}^0e^+\nu_e$, (b)$D^+\rightarrow
\bar{K}^{\star 0}e^+\nu_e$, (c) $D^+\rightarrow \pi^0 e^+\nu_e$,
(d)$D^+\rightarrow \rho^0 e^+\nu_e$, (e)$D^+\rightarrow \omega
e^+\nu_e$. The arrows in (e) show the signal region. The background
(in dashed lines) is visible only in (c) and (d).}
\label{dplus-sl}
\end{figure}

Determination of absolute branching fractions for $D$ semileptonic
decays constitute important measurements. Assuming $|V_{cx}|$ to be
known, they determine form factor normalization. Conversely, if the
form factors are known independently, say, from the lattice QCD
calculations, these branching determine the relevant $CKM$ matrix
elements. By comparing the inclusive branching fractions of the
$D^+$ and $D^0$ mesons  with the sum of the measured exclusive
branching fractions, one can determine whether there are
semileptonic decay modes as yet unobserved.
%
\begin{table*}\caption{Summary of recent absolute branching
fraction measurements of exclusive $D^+$ and $D^0$ semileptonic
decays. When only the CLEO-c absolute number is available, no
average number is provided.}\label{tab:brsemil}\begin{center}
\begin{tabular}{lccc}\hline
Decay mode & {\cal B}(\%) [CLEO-c]\cite{cleoc:excl} & {\cal B}(\%)
[BES]\cite{bes:excl} & {\cal B}(\%) [PDG06 average]\\
\hline $D^0\rightarrow K^- e^+ \nu _e$ & $3.44\pm 0.10 \pm 0.10$ &
$3.82 \pm 0.40 \pm 0.27$  & $3.47 \pm 0.13$ \\
$D^0\rightarrow \pi^- e^+ \nu _e$ & $0.262\pm 0.025 \pm 0.008$ &
$0.33 \pm 0.13\pm 0.03$ & ~~ \\
$D^0\rightarrow K^{\star -} e^+ \nu _e$ & $2.16\pm 0.15 \pm 0.08$ & ~~ & ~~ \\
$D^0\rightarrow \rho^- e^+ \nu _e$ & $0.194\pm 0.039\pm
0.013$ & ~~ & ~~
\\
\hline $D^+\rightarrow \bar{K}^0
e^+\nu _e$ & $8.71 \pm 0.38 \pm 0.37$ & $8.95\pm 1.59\pm 0.67$ & $8.7\pm 0.5$ \\
$D^+\rightarrow \pi^0
e^+\nu _e$ & $0.44 \pm 0.06 \pm 0.0.03$ & ~~ & ~~ \\
$D^+\rightarrow \bar{K}^{\star 0} e^+\nu _e$ & $5.56 \pm 0.27
\pm 0.23$ & ~~ & ~~\\
$D^+\rightarrow \rho^0
e^+\nu _e$ & $0.21 \pm 0.04 \pm 0.01$ & ~~ & ~~ \\
$D^+\rightarrow \omega e^+\nu _e$ & $0.16 ^{+0.07}_{-0.06}
\pm 0.01$ & ~~ & ~~ \\ \hline
\end{tabular}\break\end{center}
\end{table*}
BES-II\cite{bes:excl} and CLEO-c\cite{cleoc:excl} have recently
presented data on exclusive semileptonic branching fractions. BES-II
results are based on 33 pb$^{-1}$; CLEO-c's results are based on the
first 57 pb$^{-1}$ data set. Both experiments use tagged samples and
select a specific final state through the kinematic variable:
\begin{equation}
U\equiv E_{miss} -|c\vec{p}_{miss}|,\nonumber
\end{equation}
where $E_{miss}$ represents the missing energy and $\vec{p}$
represents the missing momentum of the $D$ meson decaying
semileptonically. For signal events, $U$ is expected to be 0, while
other semileptonic decays peak in different regions.
Fig.~\ref{dplus-sl} shows the $U$ distribution for 5 exclusive $D^+$
decay modes reported by CLEO-c, which demonstrate that $U$
resolution is excellent, thus allowing  a full separation between
Cabibbo suppressed and Cabibbo favored modes. Table
\ref{tab:brsemil} summarizes the recent measurements from CLEO-c and
BES-II, as well world averages reported in the Review of Particle
Physics~\cite{Yao:2006px}.

Absolute branching fractions for $\Dz\to K \ell \nu$ have been
recently published by Belle \cite{Widhalm:2006wz}: they obtain
${\cal B}(\Dz\to K \ell \nu)=(3.45\pm 0.07 \pm 0.20)$\% and ${\cal
B}(\Dz\to \pi \ell \nu)=(0.255\pm 0.019 \pm 0.016)$\%.
%
CLEO-c uses the two tagging modes with lowest background
($\overline{D}^0\rightarrow K^+\pi^-$ and $D^-\rightarrow
K^+\pi^-\pi^-$) to measure the inclusive $D^0$ and $D^+$
semileptonic branching fractions~\cite{cleoc:incl}. They obtain
$${\cal B}(D^{+}\rightarrow Xe^+\nu _e) = (16.13 \pm 0.20_{\rm stat} \pm
   0.33_{\rm sys})\%,$$
   $${\cal B}(D^{0}\rightarrow Xe^+\nu _e) =  (6.46 \pm 0.17_{\rm stat}  \pm
   0.13_{\rm sys})\%.$$
The sum of the exclusive semileptonic absolute branching fraction is
${\cal B}(D^{+}\rightarrow Xe^+\nu _e)_{\rm excl}= (15.1 \pm 0.5 \pm
0.5)$\% and ${\cal B}(D^{0}\rightarrow Xe^+\nu _e)_{\rm excl}= (6.1
\pm 0.2 \pm 0.2)$\%: the measured
 exclusive modes are consistent with saturating the
inclusive widths, although there is some room left for higher
multiplicity modes. The CLEO-c data have been used in this
comparison, as they dominate the present world average: the
exclusive modes are consistent with saturating the inclusive
semileptonic branching fraction at a 41\% confidence level in the
case of the $D^+$ and 18\% confidence level in the case of the
$D^0$.

\subsection{Form Factors For The Decays $D\to\ K(\pi) \ell \nu$}

Theoretical parameterizations of semileptonic decays involve two
non-perturbative quantities parameterizing matrix element of a
single hadronic current. Traditionally, the hadronic matrix elements
for transitions to pseudoscalar hadrons are described in terms of
two form factors, $f_+(q^2)$ and $f_-(q^2)$,
\beq
\langle K (\pi) | \bar q \Gamma^\mu c | D \rangle =
f_+(q^2) P^\mu + f_-(q^2) q^\mu,
\eeq
where $P=p_D+p_{K(\pi)}$ and $q=p_D-p_{K(\pi)}$. An alternative
parameterization is also often used,
\beq
\langle K (\pi) | \bar q \Gamma^\mu c | D \rangle =
\left(P^\mu-\frac{m_D^2-m_{K(\pi)}^2}{q^2} q^\mu \right) f_+(q^2) +
\frac{m_D^2-m_{K(\pi)}^2}{q^2} q^\mu f_0(q^2),
\eeq
with $f_0(q^2)=f_+(q^2)+f_-(q^2) q^2/(m_D^2-m_{K(\pi)}^2)$.
Experimental determinations of these form factors are performed
through the study of the differential decay width $d\Gamma /dq^2$.
For cases where the lepton in the final state is an electron and has
a negligible mass with respect to the parent D, only a single form
factor, $f_+(q^2)$, contributes. The partial decay width is given by
\begin{equation}
\frac{d\Gamma(D\to K(\pi) e \nu_e)}{dq^2} =
\frac{G_F^2|V_{cq}|^2}{24\pi^3} p_{K(\pi)}^3 |f_+(q^2)|^2
\end{equation}
where $p_{K(\pi)}$ the hadron momentum in the $D$ rest frame.
Form factors have been evaluated at specific $q^2$ points in a variety of
phenomenological models~\cite{Stone:2006zy}, where the shape is
typically assumed from some model arguments. In order to restrict the
function space studied,
a dispersive representation~\cite{Boyd:1997kz} allows to place
rather general constraints on the shapes of the form factors from
their analytic properties. Particular parameterizations of the form
factors are nevertheless useful. The most common parametrization has
been a single pole form factor, where the pole is the lowest mass
resonance formed by the initial and final state hadron. For example,
in the decay $D\to \pi e \nu_e$ the dominant pole is the $D^\star$.
Now that more precise data are available, more complex
representations are investigated. One class of parameterizations
include the dominant pole form factor and approximates the
dispersion integral by a number of effective poles
\begin{equation}
f_+(q^2)=\frac{f_+(0)}{(1-\alpha)}\frac{1}{1-(q^2/m_V^2)}+
\sum_{k=1}^N \frac{\rho _k}{1-\frac{1}{\gamma_k}\frac{q^2}{m_V^2}},
\end{equation}
where $\alpha$  determines the strength of the dominant pole,
$\rho_k$ gives the strength of the $k$th term in the expansion, and
$\gamma _k=m_{V_k}^2/m_V^2$, with $m_{V_k}$ representing masses of
the higher order poles. The true form factor can be approximated to
any desired accuracy by introducing a large number of finely spaced
effective poles. In effect, it is desirable to keep the number of
terms in this expansion to a manageable number. The popular
Becirevic-Kaidalov (BK) parametrization \cite{Becirevic:1999kt} is a
simplified version of the $N=1$ truncation of this expansion.  In
general, both the $N=0$ case (simple pole) and the $N=1$ case can
provide good representation of the data if the pole masses are
allowed to be non physical. An alternative approach
\cite{Becher:2005bg} utilizes a series expansion around an arbitrary
$q^2$ value $t_0$. To achieve a convergent series, the expansion is
formulated as an analytic continuation of the form factors in the
complex $t=q^2$ plane. There is a branch cut on the real axis for
$t>(M_D+M_{K,\pi})^2$, which corresponds to a region associated with
production of states with appropriate quantum numbers. The
convergence is accelerated by mapping the whole cut region onto the
unit disk $z<1$, where $z$ is defined as
\begin{equation}
z(q^2,t_0)=\frac{\sqrt{t_+-q^2}-\sqrt{t_+-t_0}}{\sqrt{t_+-q^2}+\sqrt{t_+-t_0}},
\end{equation}
where $t_\pm =(M_D\pm M_{K,\pi})^2$ and $t_0$ is the arbitrary $q^2$
value that maps onto $z$ =0. The form factors are then expressed as
\cite{Becher:2005bg}
\begin{equation}
f_+(q^2)=\frac{1}{P(q^2)\Phi(q^2,t_0)}\sum_{k=0}^\infty \alpha
_k(t_0)[z(q^2,t_0)]^k
\end{equation}
with $P(q^2)=z(q^2,m_V^2)$, which accounts for the pole in the
form factor at $q^2=m_V^2$. The physical observables are not expected
to depend on $\Phi(q^2,t_0)$, which can be any analytical function, or $t_0$.

\begin{figure}
\epsfxsize 25pc         %
\centerline{\epsfbox{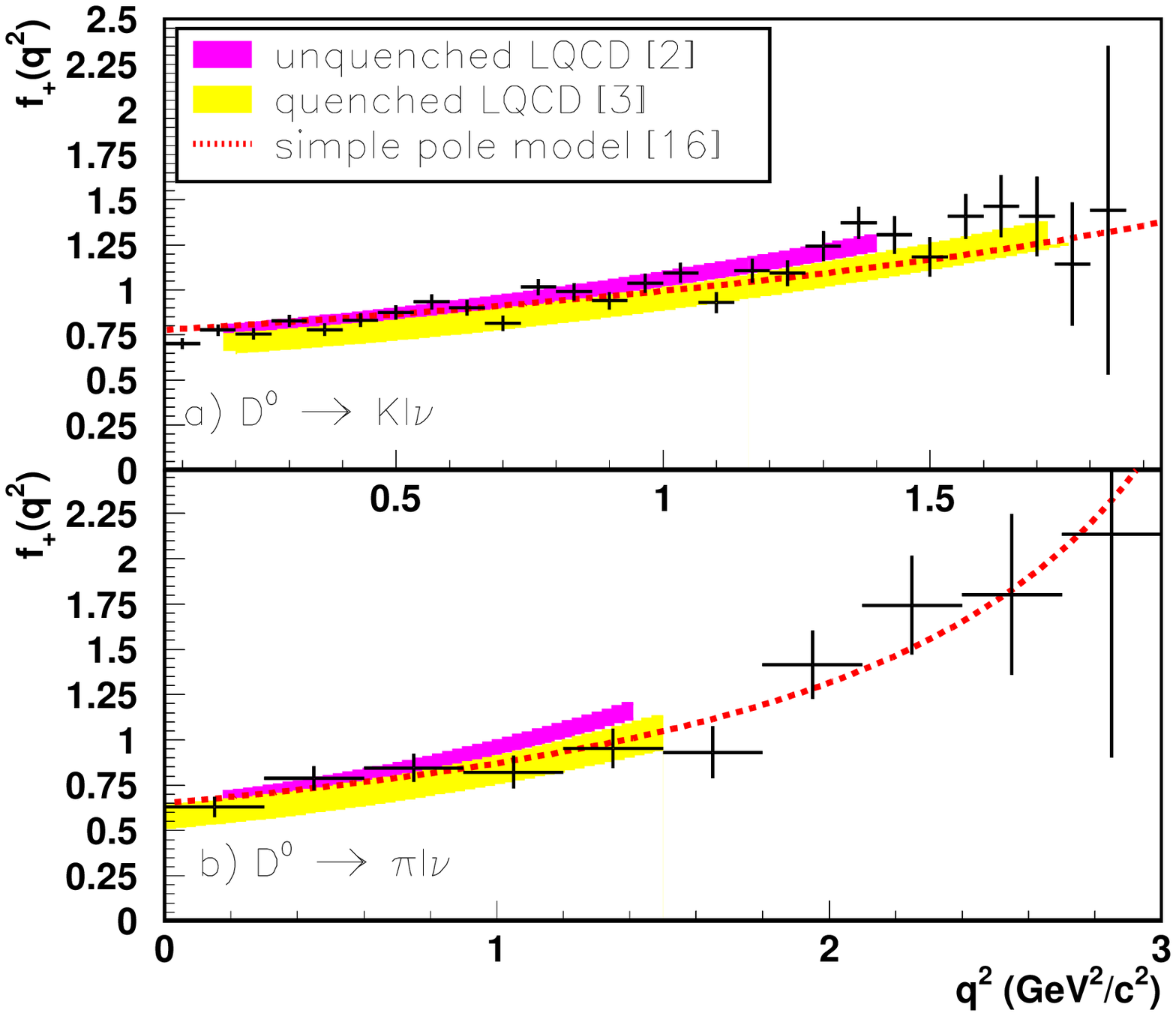}}
\caption{Belle from factors for (a) $\Dz \to K^-\ell ^+\nu$ in $q^2$
bins of 0.067 GeV$^2$ and (b) $\Dz \to \pi ^- \ell ^+ \nu$ in $q^2$
bins of 0.3 GeV$^2$. Overlaid are the predictions of the simple pole
model using the physical pole masses (dashed) and a quenched
(yellow) and unquenched (purple) LQCD calculations. The shaded bands
reflect the theoretical uncertainties and are shown in the $q^2$
ranges for which calculations are reported. \cite{Widhalm:2006wz}.}
\label{shape}
\end{figure}

Unquenched lattice $QCD$ calculations for $D\rightarrow K\ell
\bar{\nu}$ and $D\rightarrow \pi \ell \nu$ have recently been
reported~\cite{Aubin:2004ej}. The chiral extrapolation is performed
at fixed $E ={v}\cdot {p}_{K(\pi)}$. The lattice ``data points'' are fitted to the BK
parametrization~\cite{Becirevic:1999kt}
\begin{eqnarray}\label{eq:BK}
f_+(q^2)  &=&  \frac{f_+(0)}{(1-\tilde{q}^2)(1-\alpha\tilde{q}^2)},\\
\nonumber
f_0(q^2)  &=&  \frac{f_+(0)}{1-\tilde{q}^2/\beta},
\end{eqnarray}
where $\tilde{q}^2=q^2/m_{D_x^{*}}^2$, and $\alpha$ and
$\beta$ are fit parameters. The fitted parameters are shown in
Table~\ref{tab:results}.
%
%
\begin{table}
\caption{Fit parameters in Eq.~(8), decay rates and $CKM$ matrix
elements. The first errors are statistical; the second
systematic.$^{24}$} \label{tab:results}
\begin{center}
\begin{tabular}{cccc}
\hline $P$ & $F$& $\alpha$ & $\beta$ \\ \hline $\pi$ &0.64(3)(6)
&0.44(4)(7) &1.41(6) (13) \\
 $K$  &0.73(3)(7) &0.50(4)(7) &1.31(7)(13) \\
\hline
\end{tabular}
\end{center}
\end{table}
The FOCUS experiment~\cite{focus:ff} was the first to perform a
non-parametric measurement of the shape of the form factor in
$D\rightarrow K \mu \nu _{\mu}$ \cite{Link:2004dh}.  CLEO-c
\cite{Pavlunin:2005br}, Belle \cite{Widhalm:2006wz}, and BaBar
\cite{Aubert:2007wg}  reported similar analyses. Fig.~\ref{shape}
shows the lattice QCD predictions for $D\rightarrow K\ell \nu$ and
$D\rightarrow K\ell \nu$ with the Belle data points superimposed.
Table \ref{fit-data} summarizes the experimental form factor fits
compared to the lattice QCD predictions.
\begin{table}
\caption{Measured shape parameter $\alpha$ compared to lattice QCD
predictions.}\label{fit-data}
\begin{center}
\begin{tabular}{ll}\hline
\multicolumn{2}{c}{$\alpha(D^0\rightarrow K\ell \nu)$}\\\hline
Lattice QCD~\cite{Aubin:2004ej}& $0.5\pm 0.04\pm 0.07$ \\
FOCUS~\cite{Link:2004dh} & $0.28\pm 0.08\pm 0.07$\\
CLEOIII~\cite{Huang:2004fra} & $0.36\pm 0.10^{+0.03}_{-0.07}$\\
Belle~\cite{Widhalm:2006wz}& $0.52\pm 0.08\pm 0.06 $\\
BaBar~\cite{Aubert:2007wg}& $0.38\pm 0.02 \pm 0.03$ \\\hline
\multicolumn{2}{c}{$\alpha(D^0\rightarrow \pi\ell \nu$)}\\\hline
Lattice QCD~\cite{Aubin:2004ej}& $0.44\pm 0.04\pm 0.07$\\
CLEO III~\cite{Huang:2004fra} & $0.37^{+0.20}_{-0.31}\pm 0.15$\\
Belle~\cite{Widhalm:2006wz}& $0.10\pm 0.21 \pm 0.10$
\\\hline
\end{tabular}
\end{center}
\end{table}
By combining the information of the measured leptonic and
semileptonic width, a ratio independent of $|V_{cd}|$ can be
evaluated, which can serve as a check of the theoretical calculations.
For instance, assuming isospin symmetry, i.e.  $\Gamma(D\rightarrow \pi e^+\nu
_e)=\Gamma(D^0\rightarrow \pi^- e^+\nu _e)=2\Gamma(D^+\rightarrow
\pi^0 e^+\nu _e)$, a ratio
$$R\equiv\sqrt{\Gamma(D^+\rightarrow \mu \nu
_{\mu})/\Gamma(D\rightarrow \pi e^+\nu _e)}$$ can be formed. Using
the recent unquenched lattice QCD
calculations~\cite{Aubin:2005ar,Aubin:2004ej}, this ratio can be
computed to be
\begin{equation}
R^{th}_{sl}=\sqrt{\frac{\Gamma^{th}(D^+\rightarrow \mu\nu
_{\mu})}{\Gamma^{th}(D\rightarrow \pi e\nu _{e})}}=0.212\pm
0.028,\nonumber
\end{equation}
The quoted error is evaluated through a careful study of the theory
statistical and systematic uncertainties, assuming Gaussian errors.
The corresponding ``experimental'' ration can be calculated using the
CLEO-c $f_D$ and isospin averaged $\Gamma (D\rightarrow \pi
e^+\nu_e)$,
\begin{equation}
R^{exp}_{sl}=\sqrt{ \frac {\Gamma^{exp}(D^+\rightarrow\mu\nu)}
{\Gamma^{exp}(D\rightarrow\pi e \nu _e)}}= 0.249\pm 0.022.\nonumber
\end{equation}
The theoretical calculations and data are consistent at 28\% confidence level.

\subsection{The $CKM$ Matrix}

An important goal of the next generation of precision experiments is
to perform direct measurements of each individual parameter. This
will enable us to perform additional unitarity checks with precision
similar to that achieved currently with the first row~\cite{vcd}.
With the help of the unitarity constraints, charm-quark-related $V_{cd}$ and
$V_{cs}$ are now determined with rather high precision~\cite{Yao:2006px}. The most
recent results from LEP II, using the $W\rightarrow \ell \nu$
branching fraction, and additional inputs from other $CKM$ parameter
measurement is $V_{cs}=0.976\pm 0.014$~\cite{lep-win05}. The
unitarity constraint implies $V_{cd}\sim V_{us}= 0.2227\pm
0.0017$~\cite{vcd}.

CLEO-c \cite{:2007se} has extracted $|V_{cd}|$ and $|V_{cs}|$ by
combining the $|V_{cq}|f_+(0)$ results from the three parameter
series expansion fit~\cite{Becher:2005bg} with the unquenched
lattice QCD predictions for $f_+(0)$~\cite{Aubin:2004ej} to obtain
\begin{eqnarray}
|V_{cs}|  &=&  1.015 \pm 0.010 \pm 0.011 \pm 0.106 \nonumber
\\
|V_{cd}| &=& 0.217 \pm 0.009 \pm 0.004 \pm 0.023 \nonumber
\end{eqnarray}
The first two errors are experimental, statistical and systematic,
while the last errors are theoretical, dominated by the
discretization uncertainties in the lattice QCD charm quark action,
which should be improved in the near future. It will be interesting
to see a unitarity check performed on the second row of he $CKM$
matrix element once these errors are further reduced.

\subsection{Form Factors in Semileptonic $D\to V \ell \nu$ Decays}

The structure of the hadronic current in semileptonic decays
including vector mesons in the final state is more complex,
involving four independent form factors, $V$, $A_0$, $A_1$, and $A_2$,
\bea
\langle K^*(\rho) | \bar q \gamma_\mu c | D \rangle &=&
2\frac{V(q^2)}{m_D+m_{K^*(\rho)}} \epsilon_{\mu\nu\alpha\beta}
p_D^\nu p_{K^*(\rho)}^\alpha \epsilon^{*\beta},
\nonumber \\
\langle K^*(\rho) | \bar q \gamma_\mu \gamma_5 c | D \rangle &=&
i \left(m_D + m_{K^*(\rho)}\right)
\left(\epsilon^*_\mu - \frac{\epsilon^*\cdot q}{q^2} q_\mu \right) A_1(q^2)
\\
&-& i \frac{\epsilon^*\cdot q}{m_D+m_{K^*(\rho)}}
\left(P^\mu-\frac{m_D^2-m_{K(\pi)}^2}{q^2} q^\mu \right) A_2(q^2)
\nonumber \\
&+& 2i m_D \frac{\epsilon^*\cdot q}{q^2} q_\mu ~A_0(q^2),
\nonumber
\eea
where $\epsilon^*$ is a polarization of the final state meson.
The vector form factor $V$ is dominated by vector meson resonance
exchanges, $A_0$ is dominated by pseudoscalar meson resonance
exchanges,  $A_1$ and $A_2$ are dominated by axial meson resonance
exchanges. Generally a single pole form factor is assumed, both in
experimental studies and in theoretical calculations of the
normalization of the form factors. The FOCUS experiment has
developed an interesting technique that extends its non-parametric
determination of the form factors in $\Dz\to K^-\mu^+\nu_{\mu}$ to
$\Dp\to K^- \pi^+e^+ \nu_{e}$~\cite{Link:2005dp}, later adopted also
by the CLEO-c experiment~\cite{Shepherd:2006tw}. This method allows
model independent determinations of the form factors and thus
provides a better check for theoretical calculations. For example,
Fajfer and Kamenic have studied these decays by including
contributions of charm meson resonances beyond the simple pole
\cite{Fajfer:2006uy} and have found that including two poles in the
vector form factor improves the agreement between their predictions
and experimental results.

The FOCUS experiment \cite{Link:2002ev} has reported evidence for
the presence of a small even $K^-\pi^+$ amplitude interfering with
the dominant $\bar{K}^{\star 0}$ component in the decay $\Dp\to K^-
\pi^+\mu^+ \nu_{\mu}$. CLEO-c \cite{Shepherd:2006tw} has seen the
same effect in $\Dp\to K^- \pi^+e^+ \nu_{e}$. This observation opens
up new areas of investigation in exclusive charm semileptonic
decays, namely the investigation of light quark spectroscopy. For
example, it would be interesting to verify whether this broad s-wave
resonance can be identified with the $\kappa$ seen in $\Dp$ Dalitz
plot analyses~\cite{Bianco:2003vb}. It will be interesting to search
for similar interference phenomena in $\Ds$ semileptonic decays.

\section{RARE AND RADIATIVE DECAYS}\label{Rare}

\subsection{Theoretical Motivation}

Rare charm decays hold great potential to be a sensitive probe of NP.
Among all rare charm transitions, the
most interesting are the decays that are associated with $\Delta C=1$ flavor-changing
neutral currents (FCNC), i.e. transitions that
change charm quark quantum number by one unit while conserving the electrical charge
of participating quarks. Examples of
such transitions include (a) rare radiative decays mediated by $c \to u \gamma$ or
$c \to u \gamma \gamma$ quark currents, or
(b) rare leptonic and semileptonic decays mediated by  $c \to u \ell \bar\ell$ quark
currents. Here $\ell$ could either be a charged lepton
such as $e$ or $\mu$ or a neutrino $\nu$. In addition, fully non-leptonic FCNC,
such as $c \to u g$ or $c \to u q \bar q$ are possible. We shall discuss them in
Section~\ref{Hadronic}.

In the SM, where FCNC cannot occur at the tree level, this is
usually associated with large contribution of top quark to one-loop
electroweak diagrams due to the Glashow-Iliopoulos-Maiani (GIM)
mechanism~\cite{Glashow:1970gm}. This assures that the bottom-type
FCNC decay is dominated by the short-distance contributions and
therefore is reliably computable. It has become evident that this
situation is not realized in charm decays due to relatively small
mass of the bottom quark and significant hadronic dynamical effects
in the region of charmed hadron mass. This leads to overwhelming
long-distance contributions and decreased reliability of theoretical
predictions. Indeed, model-dependent evaluations of long-distance
effects are possible~\cite{Burdman:1995te,Burdman:2001tf}, which can
be used to judge relative importance of long- and short-distance
physics.

These facts can constitute a problem for proper interpretation of
new physics effects in FCNC processes. In addition, constraints on
the strength of new interactions can be unambiguously placed only if
the SM contributions are significantly smaller than the
experimentally placed bound on a branching ratio.

\subsubsection{Inclusive and exclusive radiative decays $c \to u \gamma$.}

Since rare radiative decays are two-body-decays, a branching ratio
for exclusive or inclusive transitions is the primary observable.
Thus, one has to evaluate relative NP/SM contribution for each model
of NP. Only if the SM contribution, even dominated by the LD
physics, is seen to be much smaller than current experimental bounds
and possible NP contributions, such measurements can be useful in
constraining NP models. Current theoretical estimates put decays
rates of $\Dz\to\rho\gamma$ at the level of $(0.1\div
0.5)\times10^{-5}$ and $\Dz\to\phi\gamma$ at $(0.1\div
3.4)\times10^{-5}$. Currently, the decay $\Dz \to \phi\gamma$ has
been measured to be $(2.6^{+0.70}_{-0.61}{}^{+0.15}_{-0.17})\times
10^{-5}$ \cite{tajima:2004}, and experimental constraints on other
radiative decays are of the order of $10^{-4}$~\cite{Yao:2006px}. As
the experimental bounds for radiative decays are pushed towards the
SM theoretical estimates, these decays become less and less suitable
to provide unambiguous constraints on New Physics
models~\cite{Burdman:2001tf}.

In the SM, the radiative charm decays occur via the operators of the type
$O_7=(e/16\pi^2) m_c (\overline u \sigma_{\mu\nu} P_R c) F^{\mu\nu}$.
In total, renormalization group running of perturbative QCD requires
a complete set of ten operators to describe this transition~\cite{Burdman:1995te}.
Note that due to the chiral structure of the SM, the contribution of a similar operator
$O_7^\prime=(e/16\pi^2) m_u (\overline u \sigma_{\mu\nu} P_L c) F^{\mu\nu}$
is suppressed by a small factor $m_u/m_c$. Such suppression is not universal
and is in fact absent in some models of NP, including SUSY. Thus, measurement of
polarization of the final state photon can in principle be a nice probe of
NP.

\subsubsection{Rare Decays $D\to X_u \ell^+ \ell^-$.}

\begin{figure}
\epsfxsize 25pc \centerline{\rotatebox{90}{\epsfbox{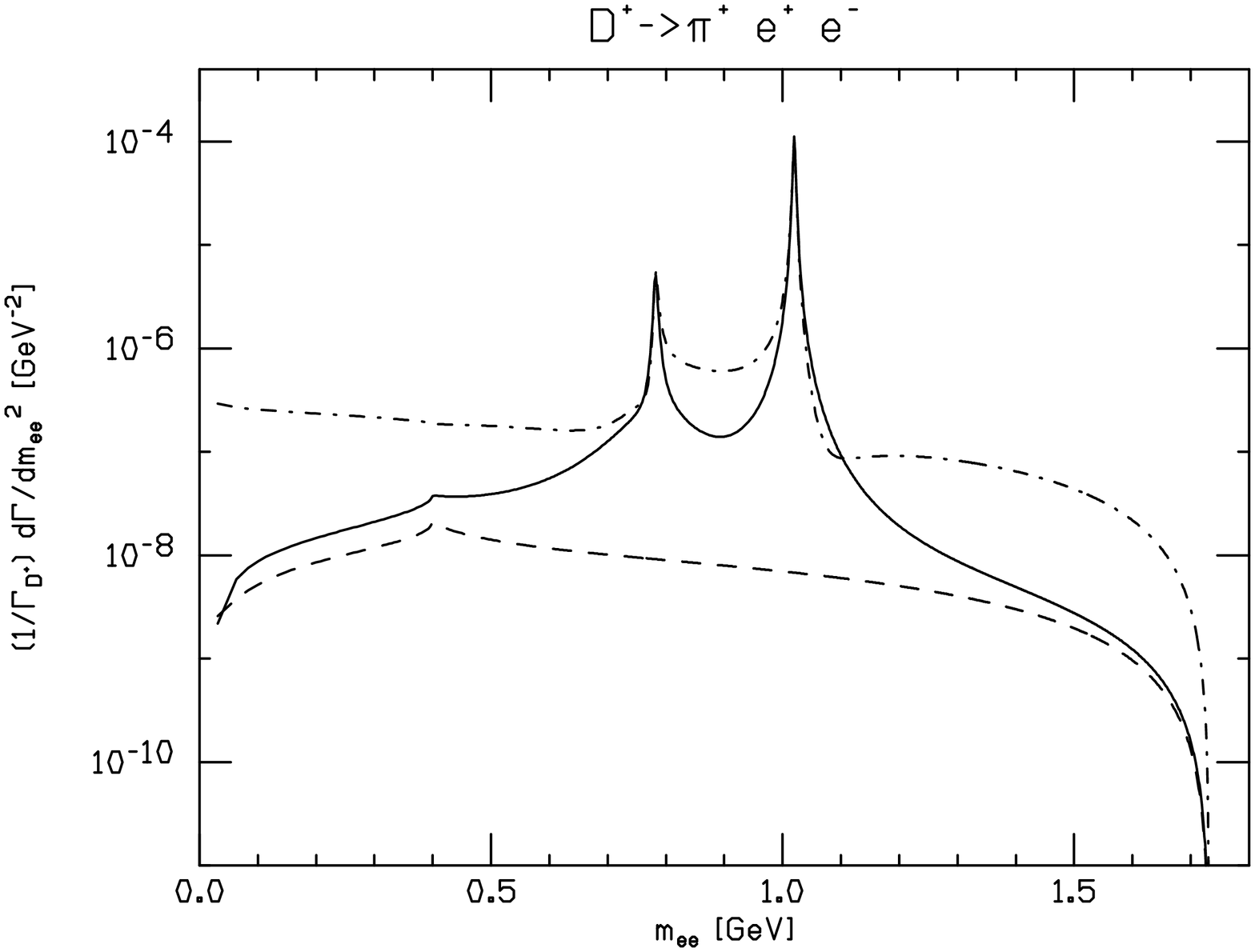}}}
 \caption{The
dilepton mass distribution for $D^+\to\pi^+ e^+e^-$ (normalized to
$\Gamma_{D^+}$) in the MSSM with nonuniversal soft breaking effects.
The solid line is the SM. (I) $M_{\tilde g}=M_{\tilde q}=250$~GeV;
(II) $M_{\tilde g}=2\,M_{\tilde q}=500$~GeV; (III) $M_{\tilde
g}=M_{\tilde q}=1000$~GeV; (IV) $M_{\tilde g}=(1/2)\,M_{\tilde
q}=250$~GeV.  Curves are from Ref. ~\cite{Burdman:2001tf}}
\label{fig:mee}
\end{figure}
Decays of the type $c\to u \ell^+\ell ^-$ may allow a better
separation of SM and NP effects. The simplest possible decay that is
generated by this current is $\Dz\to \ell^+\ell ^-$. Decays of this
type are helicity-suppressed, with decay rates proportional to the
masses squared of the final state leptons. This makes decays $\Dz
\to \epem$ prohibitively small. Even the decay $\Dz\to\mup\mum$ is
quite small. A calculation of short distance SM effects predicts a
branching fraction of about $10^{-18}$~\cite{Burdman:2001tf}. Long
distance contributions bring the predicted branching fraction to an
excess of $10^{-13}$, more precisely $2.6\times 10^{-5}\ {\cal
B}(\Dz\to\gamma\gamma)$~\cite{Burdman:2001tf}. Thus, decays of this
type provide almost background-free constraints on NP models. For
example, R-parity-violating SUSY contributions are predicted at the
level of $3\times10^{-6}$ for some region of SUSY parameter space.

In that sense, three body decays are more suitable for experimental
studies, as they do not receive the above-mentioned helicity
suppression. The two modes that have been studied most extensively
are $D \to \pi\ell ^+\ell ^-$ and $D\to \rho\ell ^+\ell ^-$. These
more complex final states provide additional tools to disentangle SM
short and long distance effects and NP phenomena. Figure
\ref{fig:mee} from Ref. \cite{Burdman:2001tf} illustrates this point
with reference to the decay $\Dp\to \pi^+e^+e^-$. It shows the
predicted dilepton mass distribution normalized to $\Gamma_{D^+}$.
The solid line represents the total SM prediction, while the curves
represent predictions from a variety of minimal supersymmetric
models. It is clear that for dilepton masses close to vector meson
resonances such as $\rho$ or $\phi$ there is no sensitivity to new
physics contributions, however there are regions where NP effects
are unambiguous. In particular, the region of low $M_{\epm}$ is of
great interest. Similar considerations apply to $D\to
\rho\ell^+\ell^-$, where additional information is provided by the
lepton forward-backward asymmetry
\begin{equation}
A_{FB}(q^2)\equiv \frac{\int_0^1 d\Gamma/(dq^2dx) dx-\int_{-1}^0
d\Gamma/(dq^2dx) dx }{d\Gamma/dq^2}
\end{equation}
where $x = \cos{\theta}$. Here $\theta$ is the angle between the
$\ell ^+$ and the $D$ meson in the $D$ rest frame. In the SM
$A_{FB}(q^2)$ is negligibly small for all values of $q^2$.

\subsection{Experimental Information}
A rare $D$ radiative decay has recently been seen by
Belle~\cite{tajima:2004} with the reported branching ratio ${\cal
B}(\Dz\to \phi\gamma) =(2.6^{+0.70}_{-0.61}{}^{+0.15}_{-0.17})\times
10^{-5}$. This branching fraction is measured by studying
simultaneously the decays $\Dz \to\phi \gamma$, $\Dz\to \phi\piz$,
and $\Dz\to\phi \eta$, as the last two modes, with higher branching
fractions, induce some peaking background. $\Dz \to\phi \gamma$ is
dominated by long distance effects, and the branching fraction is at
the level expected from theoretical estimates.

The most stringent limits on the di-lepton channels have been
obtained recently by CDF~\cite{cdf:2mu}, who study $\Dz \to
\mu^+\mu^-$ and BaBar~\cite{Aubert:2006aka} who study both $\Dz \to
\mu^+\mu^-$ and $\Dz \to e^+e^-$. Both experiments use a $D^\star$
tagged sample and normalize their results with respect to $\Dz\to
\pi^+\pi^-$.

The channel $c\to u \ell^+\ell^-$ has been studied by
CLEO-c\cite{He:2005iz}, BaBar \cite{Aubert:2006aka}, and D0
\cite{:2007kg}.  The CLEO-c study focuses on $\Dp \to \pi ^+ e^+e^-$
because of their excellent sensitivity to di-electron final states,
while BaBar studies both di-muon and di-electron final states of
$\Dp$, $\Dsp$ and $\Lambda _c^+$. D0 has very recently reported
results on the $\Dp \to \pip\mup\mum$ final state. All three
experiments start by measuring $\Dp\to \pip\phi\to \ell
^+\ell^-\pip$ to establish the order of magnitude of long distance
effects. CLEO-c finds two events with an expected background of 0.02
events, BaBar finds 19 events over a background of 40 events, and D0
finds 115 events over a background of 850 events. They find the
branching fractions:
\begin{eqnarray}
{\cal B}(\Dp\to \phi \pi^+ \to e^+e^- \pi^+) &=&  (2.8\pm 1.9\pm
0.2)\times 10^{-6} ~~\left[{ \cite{He:2005iz}}\right]
\nonumber \\
&=&  (2.7^{+3.6}_{-1.8})\times 10^{-6} ~~\left[
\cite{Aubert:2006aka}\right],
\end{eqnarray}
and
\begin{equation}
{\cal B}(\Dp\to \phi \pi^+ \to \mu^+\mu^-\pi^+)  = (1.8\pm 0.5\pm
0.6)\times 10^{-6} ~~\left[{\rm \cite{:2007kg}}\right].
\end{equation}
These experiments establish also 90 \% confidence level upper limits
on the short distance components of these branching fractions. Their
results, compared with a representative sample of theoretical
expectations based on NP scenarios are summarized in Table
~\ref{drare:tab}. These data show that experiments are reaching a
sufficient sensitivity to limit the NP parameter space. The
availability of higher statistics data samples from $\epm$
experiments and collider experiments, as LHCb, when data taking
starts, bear the promise of more stringent tests in the near future.
\begin{table}
\caption{ Representative predictions for flavor changing neutral
current charm decays and experimental upper
limits.\label{drare:tab}}
\begin{tabular}{lccc}
\hline Decay Mode & LD $(\times 10^6$)&  MSSMR ($\times 10^6$)
& Experiment (90 \% CL ul $\times 10^6$)
\\\hline $\Dp\to \pip\epm$& 2.0 \cite{Fajfer:2007dy}&
$0.21\cite{Fajfer:2007dy}-2.0\cite{Burdman:2001tf}$ & 7.4\cite{He:2005iz}  \\
~~& ~~& ~~ & 11.2 \cite{Aubert:2006aka}\\ \hline $\Dp\to
\pip\mup\mum$ & $1.9$ &
$6.5\cite{Fajfer:2007dy}-15\cite{Burdman:2001tf}$ &
24.4\cite{Aubert:2006aka}
\\
~~& ~~& ~~ &3.9\cite{:2007kg}\\\hline $\Dp\to \pip\mup\ep$ & 0 &
30\cite{Burdman:2001tf} & 10.8\cite{Aubert:2006aka}
\\\hline
$\Dz\to\epm$ &$1.0\times 10^{-17}$ & $1.0\times 10^{-4}$& $1.2$\cite{Aubert:2004bs}\\
$\Dz\to\mup\mum$ & $3.0\times 10^{-7}$ & $3.5 $   &  1.3\cite{Aubert:2004bs}\\
~~& ~~& ~~ & 2.5 \cite{Acosta:2003ag}\\ \hline
$\Dz\to \rho ^0\epm$ &
$1.8$ & 5.1 & 100.0 \cite{Freyberger:1996it}
\\
\\\hline
\end{tabular}
\end{table}

\section{HADRONIC DECAYS}\label{Hadronic}

Hadronic decays are interesting for several reasons.  Absolute
measurements of $D$ meson branching fractions affect our knowledge
of several $D$ and $B$ meson decays, from which $CKM$ parameters are
extracted. Multi-body final states provide information on light
quark spectroscopy. Nonleptonic decays of charmed hadrons provide
needed information~\cite{Cavoto:2006um} which helps in
determinations of $CKM$ angles $\beta$~\cite{Aubert:2004cp} and
$\gamma$ \cite{Giri:2003ty} in $B$ decays and, can also help in
determination of $\DzDzb$ mixing parameters free from knowledge of
hadronic strong phases~\cite{Abe:2007rd}.

\subsection{Theoretical Considerations}

Theoretical description of fully hadronic decays is significantly
more complicated than leptonic or  semileptonic ones, even though
relevant effective Hamiltonians look quite similar to
Eq.~(\ref{Hsl}). Charmed nonleptonic decays are usually classified
by the degree of $CKM$ suppression.  Least suppressed, where the
quark level transitions are $c\to su\bar d$ are labeled ``Cabibbo
favored" (CF) decays and governed by
\bea\label{Hcf}
{\cal H}_{\tiny CF} &=& \frac{G_F}{\sqrt{2}} V_{ud} V^*_{cs}
\left[
C_1(\mu) {\cal O}_1 + C_2(\mu) {\cal O}_2 \right] + {\rm h.c},
\nonumber \\
{\cal O}_1 &=& \left(\overline{s}_i \Gamma_\mu c_i\right)
\left(\overline{u}_k \Gamma^\mu d_k\right), \quad
{\cal O}_2 = \left(\overline{s}_i \Gamma_\mu c_k\right)
\left(\overline{u}_k \Gamma^\mu d_i\right)
\eea
where $C_n(\mu)$ are the Wilson coefficients obtained by
perturbative QCD running from $M_W$ scale to the scale $\mu$
relevant for hadronic decay, and the Latin indices denote quark
color.

The ``Cabibbo suppressed" (CS) transitions are driven by $c\to du\bar d$ or $c\to
su\bar s$ quark processes. Due to the presence of the quark-antiquark pair of the
same flavor in the final state, the effective Hamiltonian takes much more elaborate
form,
\bea\label{Hcs}
{\cal H}_{\tiny CS} &=& \frac{G_F}{\sqrt{2}} \sum_{q=s,d} V_{uq} V^*_{cq}
\left[
C_1(\mu) {\cal O}_1^q + C_2(\mu) {\cal O}_2^q \right]
\nonumber \\
&-& \frac{G_F}{\sqrt{2}} V_{ub} V^*_{cb} \sum_{n=3}^6 C_n (\mu) {\cal O} +  {\rm h.c},
\nonumber \\
{\cal O}_1 &=& \left(\overline{q}_i \Gamma_\mu c_i\right)
\left(\overline{u}_k \Gamma^\mu q_k\right), \quad
{\cal O}_2 = \left(\overline{q}_i \Gamma_\mu c_k\right)
\left(\overline{u}_k \Gamma^\mu q_i\right),
\nonumber
\eea
where $q=d,~s$, and ${\cal O}_{3-6}$ are the so-called ``penguin'' operators of the
type $(\overline{u} c)_{V-A}\sum_q (\overline{q} q)_{V\pm A}$
(see, e.g. Ref.~\cite{Buccella:1994nf}).

The ``Doubly Cabibbo suppressed" (DCS) decay is the one in which $c\to du\bar s$
quark transition drives the decay. The effective Hamiltonian for
DCS decay can be obtained from Eq.~(\ref{Hcf}) by interchanging
$s\leftrightarrow d$.

Calculations of hadronic decay rates governed by these transitions are quite
complicated and model-dependent. Most often, simplified assumptions, such as
factorization~\cite{Bauer:1986bm,Buras:1985xv} are used to estimate the needed
branching ratios. Some dynamical approaches, such as QCD sum rules, have been
used to justify those assumptions~\cite{Blok:1992hw}. Charmed mesons
populate the energy range where non-perturbative quark dynamics is active.
This leads to resonance effects that affect the phases of nonleptonic decay
amplitudes~\cite{Falk:1999ts}. Finally, standard methods of flavor $SU(3)$ can
be used in studies of non-leptonic D-meson decays~\cite{Savage:1991wu}.

\subsection{Branching Fraction Measurements}

CLEO-c use tagged samples to obtain precise values for absolute
hadronic branching fractions for $\Dz$ and $\Dp$
\cite{Dobbs:2007zt}, and for $\Ds$ \cite{Alexander:2008cq}. Their
``double-tag technique'' is similar to the one developed by Mark III
\cite{Baltrusaitis:1985iw}. From data at the $\psi(3770)$ they use
three $\Dz$ decay modes ($\Dz\to \Km\pip$, $\Dz\to \Km\pip\piz$,
$\Dz\to \Km\pip\pim\pip$), and 6 $\Dp$ modes ($\Dp\to \Km\pip\pip$,
$\Dp\to\Km\pip\pip\piz$, $\Dp\to \KS\pip$, $\Dp\to\KS\pip\piz$,
$\Dp\to \KS\pip\pim\pip$, $\Dp\to \Kp\Km\pip$. Single and double tag
yields are used to extract the branching fractions and $D\bar{D}$
yields  from a combined fit \cite{Sun:2005ip} to all the measured
yields. This powerful technique, combined with careful efficiency
studies based on data, dominates the present world
averages~\cite{Yao:2006px}.  Corrections for final state radiation
are included in these branching fractions. The large numbers of
$D\bar D$ pairs, $N_{\Dz\Dzbar}=(1.031\pm 0.017)\times10^6$ and
$N_{\Dp\Dm}=(0.819\pm 0.012)\times 10^6$ assure measurements at the
3\% level, limited by systematic uncertainties. They apply a similar
technique to derive absolute $\Ds$ branching fractions from data at
a center-of-mass energy of $\sim 4170$ MeV \cite{Alexander:2008cq}.
Here the dominant final state is $\Ds\Dsstar$, thus the analysis is
more complex because of the $\gamma$ from the $\Dsstar$.

$B$ factories use $D$ samples produced inclusively in $B$ meson
decays, or partially reconstructed samples of $D$ or $D_s$ recoiling
against a fully reconstructed charmed meson as normalization.
Absolute branching fraction measurements for hadronic decays of
$\Dz$, $\Dp$, and $\Ds$ are summarized in Table~\ref{tab:refmodes},
which include new absolute BF's for $\Dz\to\Km\pip$, reported by
\babar\ \cite{Aubert:2007wn} with approximately a 2\% uncertainty,
and from Belle \cite{Abe:2007jz} who measured the corresponding
quantity for $\Ds\to\Kp\Km\pip$ with a precision of $\sim 14$\%.
\begin{center}
\begin{table}%
\def~{\hphantom{0}}
\caption{Recent absolute branching fraction data.  For CLEO-c
results, the uncertainty
 due to radiative corrections has been absorbed into the
 systematic uncertainty.}\label{tab:refmodes}
\begin{center}
\begin{tabular}{@{}lcc@{}}%
\toprule Mode                & Absolute BF (\%)
                    & PDG\cite{Yao:2006px} (\%) \\
\colrule $\Dz\to\Km\pip$     & $4.007\pm 0.037\pm 0.070$
                      \cite{Aubert:2007wn}
                    & $3.82\pm 0.07$               \\
                    & $3.891\pm 0.035\pm 0.069$
                      \cite{Dobbs:2007zt}     \\
$\Dz\to\Km\pip\piz$ & $14.57\pm 01.2\pm 0.069$ \cite{Dobbs:2007zt}  & \\
$\Dz\to\Km\pip\pip\pim$ & $8.30\pm 0.07 \pm 0.38 $ \cite{Dobbs:2007zt} & \\
                      \hline
$D^+\to\Kp\pip\pip$ & $9.14\pm 0.10\pm 0.17$
                      \cite{Dobbs:2007zt}
                    & $9.51\pm 0.34$               \\
$D^+\to\Kp\pip\pip\piz$ & $5.98\pm 0.08\pm 0.01$
                      \cite{Dobbs:2007zt}
                    & $6.00\pm 0.28$               \\
$D^+\to\KS\pip$ & $1.539\pm 0.022\pm 0.038$
                      \cite{Dobbs:2007zt}
                    & $1.47\pm 0.06 $               \\
$D^+\to \KS\pip\piz$ & $7.05\pm 0.09\pm 0.25$
                      \cite{Dobbs:2007zt}
                    & $7.0 \pm 0.5$               \\
$D^+\to \KS\pip\pip\pim$ & $3.149\pm 0.046\pm 0.096$
                      \cite{Dobbs:2007zt}
                    & $3.11\pm 0.21$               \\
$D^+\to\Kp\Km\pip$ & $0.935\pm 0.017\pm 0.024$
                      \cite{Dobbs:2007zt}
                    & $1.0\pm 0.04$               \\
                    \hline
$\Ds\to\Kp\Km\pip$ & $5.50\pm 0.23\pm 0.16$ \cite{Alexander:2008cq}
                    & $5.3\pm 0.8$                 \\
                    & $4.0\pm 0.4\pm 0.4$
                      \cite{Abe:2007jz} \\
$\Ds\to\KS\Kp $ & $1.49\pm 0.07 \pm 0.05$ \cite{Alexander:2008cq}
                    & $2.2\pm 0.45$                 \\
$\Ds\to\Kp\Km\pip\piz$ & $5.62\pm 0.33\pm 0.51$
\cite{Alexander:2008cq}
                    & -               \\
$\Ds\to\pip\pim\pip$ & $1.11\pm 0.07\pm 0.04$
\cite{Alexander:2008cq}
                    & $1.22\pm 0.23$                 \\
$\Ds\to\pip\eta $ & $1.47\pm 0.12\pm 0.14$ \cite{Alexander:2008cq}
                    & $2.11\pm 0.35$                 \\
$\Ds\to\pip\etap$ & $4.02\pm 0.27\pm 0.30$ \cite{Alexander:2008cq}
                    & $4.7\pm 0.7$                 \\
\hline \botrule
\end{tabular}
\end{center}
\end{table}
\end{center}
In the \babar\ measurement \cite{Aubert:2007wn},
 $\bar B^0\to D^{*+}X\ell^-\bar\nu_{\ell}$
with $D^{*+}\to\Dz\pip_s$ are identified by partial reconstruction.
Events with a lepton $\ell^-$ and a slow pion $\pip_s$ that could
come from a $D^{*+}$ are selected by studying the reconstructed
$\nu$ invariant mass squared $M_{\nu,\hbox{meas}}^2$ inferred from
conservation of energy and momentum in the center-of-mass (CMS)
system, using the $D^\star$ 4-momentum inferred from $\pi_s$ and the
measured charged lepton's 4-momentum. The ${\cal B}$ normalization is
determined from the peak in $M_{\nu,\hbox{meas}}^2$ (centered
near 0 for real $\Dz$ events).  Backgrounds under this peak, coming from
$D^{**}$, other $B\bar B$ combinations and continuum, are estimated
from a wrong-sign lepton sample.  Uncertainties in these backgrounds
and in charged track reconstruction and particle
identification efficiencies, dominate the systematic errors.

The Belle result \cite{Abe:2007jz} uses partially reconstructed
$\epem\to D_s^{*+}D_{s1}^-(2536)$ events where
$D_s^{*+}\to\Ds\gamma$. They study two partially reconstructed
samples: in the first, the $\Ds$ is not reconstructed, but the
$D_{s1}^-(2536)$ is fully reconstructed in its  decay to $\bar
D^{*0}\Km$ and $\bar D^{*-}\KS$. The soft $\gamma$ from the
recoiling $D_s^{*+}$ is also required.  In the other normalization
sample, the $D_s$ is fully reconstructed in its $\Kp\Km\pip$ decay
mode and is combined with the $\gamma$ to form a $D_s^{*+}$. A
recoil $K$ from the $D_{s1}$ decay was also required but the
$D_{s1}^-(2536)$ was not reconstructed.  The result, obtained from
the ratio of the $\Kp\Km\pip$ signal in the first mode to the
$D_{s1}$ signal in the second, together with the (well-known)
$D^{*+}\to\Dz\pip$ and $D_s^{*+}\to\Ds\gamma$ BF's, provided the
required BF for $\Ds\to\Kp\Km\pip$.

Of special interest are the decays to $K^0$. CLEO-c
\cite{White:2007br} has recently studied both the $\KL\pi$ and
$\KS\pi$ final states. The $\KL$ are identified as a peak in the
missing mass. Effects of quantum correlations from the coherent $D$
pairs from $\psi(3770)$ decay are carefully taken into account. They
measured the asymmetries
\[  R(D) = {{\cal B}(D\to\KS\pi)-{\cal B}(D\to\KL\pi)\over
            {\cal B}(D\to\KS\pi)+{\cal B}(D\to\KL\pi)}
\]
for $D=\Dz$ and $D=\Dp$.  As pointed out in Ref.~\cite{Bigi:1994aw}
$\Dz\to K^0\piz$ involves interference between CF
and DCS modes and, since we observe $K^0$ without knowing its
strangeness, the $\KS$ and $\KL$ are related to give an asymmetry
$R(D)=2\tan\theta_C$ where $\theta_C$ is the Cabibbo angle.  Rosner
has observed \cite{Rosner:2006bw} that SU(3) flavor symmetry,
specifically $U$-spin symmetry, predicts that the ratio of
amplitudes for $\Dz\to K^0\piz$ to $\Dz\to\bar K^0\piz$ is
$\tan^2\theta_C\sim 0.054$. This leads to the prediction that the
value for $D(\Dz)$ should be $2\tan^2\theta_C \sim 0.109\pm 0.001$.
Ref.~\cite{White:2007br} finds $D(\Dz)=0.108\pm 0.025\pm 0.024$,
significantly different from zero, and in good agreement with this
prediction. There are no predictions for $R(D^+)$, measured to be
$0.022\pm 0.016\pm 0.018$, compatible with zero.

\subsubsection{Cabibbo Suppressed Hadronic Decays.}\label{sec:had_cs}

Due to $CKM$ suppression, these rates are expected to be lower by a
factor
$\rCS=|\left(V_{cs}V_{us}\right)/\left(|V_{cs}V_{ud}\right)|^2
\approx 0.05$ relative to CF rates.

Using their 281~ pb $^{-1}\ \psi(3770)$ sample, CLEO-c
\cite{Rubin:2005py} measured branching fractions for many
multi-pion, $\eta$ and $\omega$ decay modes of $\Dz$ and $D^+$
mesons. They use single tags and they extract absolute branching
fractions using the corresponding well measured CF modes for
normalization. These branching fractions range from $(1-4)\times
10^{-3}$ and are measured with a precision of about $(5-10)$\%. The
largest rates are $\Dz\to\pim\pip\piz$ ($13.2\pm 0.6\times 10^{-3}$)
for $\Dz$ and $D^+\to\pim\pip\pip$ ($3.35\pm 0.22\times 10^{-3}$)
and $D^+\to\eta\pip$ ($3.61\pm 0.36\times 10^{-3}$) for $D^+$. These
measurements represent a significant improvement on previous
knowledge, frequently being first observations. The results are
generally consistent with simple $CKM$ suppression. However, only an
upper limit is extracted for $\Dz\to 3\piz$, in spite of the large
${\cal B}$ observed for $\Dz\to\pim\pip\piz$. A possible explanation
is that the 3 pions are produced predominantly in an $I=0$ state,
inaccessible to this mode, after a $\Delta I=1/2$ transition.

The di-pion modes, $D^+\to\pip\piz$, $\Dz\to\pip\pim$ and
$\Dz\to\piz\piz$ are related by two amplitudes $A_0$ and $A_2$
corresponding, respectively, to the $S$-wave di-pion isospin $I=0$
and $I=2$ states produced
\bea
  A^{+0} &=& \sqrt{3\over 2}A_2 \quad
  A^{+-} = \sqrt{2\over 3}A_0 + \sqrt{1\over 3}A_2 \nonumber\\
  A^{00} &=& \sqrt{1\over 3}A_0 - \sqrt{2\over 3}A_2
\eea
Following the procedure outlined in Ref.~\cite{Selen:1993pt},
CLEO obtains~\cite{Rubin:2005py}
from these new results
$|A_2/A_0|=0.420\pm 0.014\pm 0.01$ and $\arg(A_2/A_0)=(86.4\pm 2.8\pm
3.3)^{\circ}$, which is rather large.

A long standing puzzle is found in the ratio ${\cal
R}(\Dz\to\Kp\Km)/{\cal R}(\Dz\to\pip\pim)=3.53\pm0.12$ where ${\cal
R}$ are BF's corrected for the 2-body phase space factor.

\babar\ \cite{Aubert:2006xn} has measured the branching fractions
\bea
  {\cal B}(\Dz\to\pip\pim\piz)=(1.493\pm 0.008\pm 0.055)\times 10^{-2}
  \nonumber\\
  {\cal B}(\Dz\to\Kp\Km\piz)=(0.334\pm 0.004\pm 0.015)\times
  10^{-2}.
  \nonumber
\eea
The corresponding ratio for decays where an extra $\piz$ is produced
gives  ${\cal M}(\Dz\to\Kp\Km\piz)/{\cal
M}(\Dz\to\pip\pim\piz)=0.678\pm 0.027$, in clear contrast to the
2-body ratio above.  This result has recently been confirmed by
Belle~\cite{Abe:2006tv}.

CLEO-c also studied $\Ds\to PP$ modes \cite{Adams:2007mx}, where
``$P$" is any pseudo-scalar meson, using their $\sqrt s=4170$ MeV
sample. The modes ($\Kp\eta$, $\Kp\eta^{\prime}$, $\pip\KS$,
$\Kp\piz$ are seen for the first time and are compared with their CF
counterparts $\pip\eta$, $\pip\eta^{\prime}$ and $\Kp\KS$. The
ratios observed are reasonably consistent with the value of
$\rCS\approx 5$\%. The decays $\Ds\to\pip\piz$ have not yet been
seen. The di-pions would be in an $S$-wave with $I=2$, and would
have to be reached through a $\Delta I=2$ transition, apparently
much suppressed.

\subsubsection{Double Cabibbo Suppressed Hadronic Decays $(c\to du\bar s)$ of $D$ Mesons.}

These decays are expected to be suppressed relative to CF modes by a
factor
$\rDCS=|\left(V_{cd}V_{us}\right)/\left(|V_{cs}V_{ud}\right)|^2
\approx 3.1\times 10^{-3}$.  For $\Dz$'s, these rates are comparable
to the mixing rate so that the two processes interfere; therefore
disentangling the two effects needs some care. This is discussed in
more detail in Section~\ref{Mixing}.

With such small branching ratios, one might wonder if DCS
transitions could be affected by NP effects. However, since the
final state is composed of quarks of different flavors, it is hard
to find a well-motivated NP model that can affect DCS transition at
an appreciable level~\cite{Bergmann:1999pm}.

The $\Dz$ DCS branching fractions measured can be related, using world averages~\cite{Yao:2006px},
to the expectations based on the value of
$\rDCS$
\[\begin{array}{lcl}
    (\Kp\pim)                              &
      (1.45\pm 0.04)\times 10^{-4}         &
      (1.18\pm 0.26)\times \rDCS  \\
    (\Kp\pim\piz)                          &
      (2.96\pm 0.19)\times 10^{-4}         &
      (2.50\pm 0.57)\times \rDCS  \\
    (\Kp\pim)                              &
      (2.49^{+0.21}_{-0.19})\times 10^{-4} &
      (2.10\pm 0.49)\times \rDCS.
  \end{array}\]

The only decay of $\Dp$, free from mixing effects, so far observed
is $\Dp\to\Kp\piz$.
\babar\ \cite{Aubert:2006sh} obtain ${\cal B}=(2.52\pm 0.47\pm
0.25\pm 0.08)\times 10^{-4}$ and CLEO-c \cite{Dytman:2006ha} obtain
${\cal B}=(2.28\pm 0.36\pm 0.15\pm 0.08)\times 10^{-4}$.  Both
experiments use the $D^+\to\Km\pip\pip$ as normalization, and the
fourth uncertainty in each case is due to this.

The decay of $\Dp\to\Kp\piz$ has been observed by \babar\ and
confirmed with a more precise measurement by CLEO-c.
Combining these measurements with the known lifetimes for $\Dz$ and
$D^+$ and the $\Dz\to \Km\pip$ BF's from ref.~\cite{Yao:2006px}, a ratio
can be formed,
\[  {\Gamma\left(D^+\to\Kp\piz\right)\over
     \Gamma\left(D^0\to\Km\pip\right)} =
     (2.44\pm 0.33)\times 10^{-3} =
     (0.79\pm 0.20)\times \rDCS \]
which is clearly compatible with the expected DCS rate. Similarly,
the ratio of the two DCS rates is
\[  {\Gamma\left(D^+\to\Kp\piz\right)\over
     \Gamma\left(D^0\to\Kp\pim\right)} = 0.66\pm 0.09.
\]
A naive spectator diagram analysis would predict a ratio of 0.5. The
difference is probably due to final state interaction effects,
although annihilation or exchange diagrams could also contribute.

\subsection{Three Body Decays}
Multi-body $D$ meson decays are a very rich source of information on
long range strong interaction effects because of the complex
interference patterns between intermediate resonances formed between
hadrons in the final states.  In addition, some of them are relevant
for \CPV\ measurements in $B$ decays \cite{Giri:2003ty} or $\DzDzb$
mixing parameters \cite{Aubert:2004cp}. A vast body of experimental
data has been studied, from fixed target experiments, CLEO (from data sets both near
the $\Upsilon(4S)$ and at 3770 MeV center-of-mass energies), \babar\
and Belle.  Huge data samples - of order $10^6$ events in many channels
- from the $B$ factories have led to the need to review models used
to fit the Dalitz plot distributions.  Many results, no doubt, await
such review and are yet to be published. Of particular concern is
the extent to which phase information, essential to the
determination of the parameters above, depends on the models assumed
for these fits.  Efforts are underway to attempt less
model-dependent approaches \cite{Meadows:2007jm,Bondar:2008hh}.

\subsubsection{Formalism for Three-Body $D$ and $\Ds$ Decays}

Decays of $D$ or $\Ds$ to three hadrons $ABC$ often proceed through
quasi 2-body modes $D\to A+f$ followed by $f\to B+C$, where $f$ is
an intermediate ``isobar" state, as outlined in
Ref.~\cite{Bauer:1986bm}. When $A,~B$ and $C$ are pseudo-scalar
hadrons, the Dalitz plot, in which the squared invariant mass of one
hadron pair is plotted against the squared invariant mass of one of
the other pairs, contains all the dynamical information. These
Dalitz plots often show intricate interference patterns between
multiple resonances that may be produced in an intermediate state.

The decay amplitude could be constructed
from a partial wave expansion in any one of the three possible
channels $f$ defined by the particle pair.  Here we choose $f=BC$.
Each wave would then be characterized by the spin ($J=L$ for
pseudo-scalar hadrons) and isospin $I$ of $f$
\bea
  {\cal A}(s,s') &=&
          \sum_I\sum_{L=0}^{\infty} M_L(p,q) F_{L,I}(s)
  \label{eq:pwa}
\eea
where $s$ and $s'$ are squared invariant masses for $BC$ (\ie\ $f$)
and for the $AC$ channel, respectively.  $F_{L,I}$ is the partial
wave decay amplitude for the system $f$.  $M$ is a tensor function
appropriate for the conservation of
total spin in the decay and depends on $L$ and the momenta
$p$ and $q$ of $B$ and $A$, respectively, that are defined in the $f$
rest frame.  The density of points on the Dalitz plot is then
proportional to $|{\cal A}|^2$.

The complex function $F_{L,I}(s)$ describes the production
and final state scattering resulting in the observed system $f$.
Two distance scales may be distinguished.  In
the first, the parent $D$ decays weakly, and hadronization occurs
making an intermediate hadron state $k$ that may differ from $f$. At
longer range, re-scattering (e.g. $KK\to\pi\pi$) occurs to make the
observed system $f$.  Thus, dropping label $L$ and $I$, $F_{L,I}(s)$
can be written
\bea
  F_f(s) = T_{fk}(s)Q_k(s)
  \label{eq:factorize}
\eea
in which $T_{fk}(s)$ is the matrix that describes hadron-hadron
scattering. When $s$ is small, $T$ can only include elastic
scattering. In this regime, therefore, {\sl in the absence of
scattering between $k$ and the recoil hadron $A$}, the Watson
theorem \cite{Watson:1952ji} requiring that the phase of $F_f(s)$
should have the same $s$-dependence as elastic scattering, should
hold. However, $Q_k(s)$ is an unknown function describing the
short-range effects and could well have an $s$-dependent phase when
strong $k-f$ scattering takes place.  In this case, the Watson
theorem would not hold.  A recent $K$-matrix fit to the $I=1/2$
$\Km\pip$ $S$-wave amplitude from $\Dp\to\Km\pip\pip$ data by the
FOCUS collaboration~\cite{Pennington:2007se} indicate that this may
be so.

\subsubsection{Analysis Methods of Three-Body $D$ and $\Ds$ Decays}

In fitting Dalitz plots, analysts have used several assumptions for
the form of $F_f(s)$.  Most often used so far is the Breit-Wigner
``isobar model" in which $F(s)$ is approximated as a linear
combination of resonant terms for each wave.  Eq.~(\ref{eq:pwa})
is approximated by a finite sum that includes only
terms for the resonant states $r$ observed in the data, {\sl no
matter in which channel they occur}.  The amplitude (\ref{eq:pwa})
is then a sum of Breit-Wigner propagators
\bea
  {\cal A}(s,s') &=& NR + \sum_r A_r \nonumber \\
        A_r      &=& c_re^{i\delta_r}
                     \frac{M_L(p,q)G_L(q)G_L(p)}{m_r^2 - s + im_r\Gamma(s,L)}
  \label{eq:isobarmodel}
\eea
with complex coefficients $c_re^{i\delta_r}$ and a constant term
$NR$,  often introduced to describe direct, non-resonant decay
to the three hadrons $A, B$ and $C$.  The parent $D$ and the resonance
have form-factors $G_L$ that depend on $L$.
Here, $m_r$ is the mass, and
$\Gamma(s,L)= \Gamma_r(m_r/\sqrt s)(p/p_r)^{2L+1}[G_L(p)/G_L(p_r)]^2$
is the mass-dependent width of the resonance $r$.  The form-factors
$G_L(q)$ and $G_L(p)$ for, respectively, the parent and the resonance
$r$, are usually assumed to take the Blatt-Weisskopf form~\cite{BlattWeiss}.
Fractions are defined, for each resonance $r$, as
\beq
  f_r = \frac{\int\int |A_r|^2 dsds'}{\int\int |NR+\sum_r A_r|^2dsds'}.
  \label{eq:isofrac}
\eeq
The sum of fractions, so defined, is not required to be unity, since
the interference terms, included in the denominator, are missing from
the numerator.  Though Eq.~(\ref{eq:isofrac}) is in standard use, a
better definition would be desirable.  In particular, it works only
to define resonant fractions from the isobar model.

Other models attempt to address the problems associated with the
description of the $S$-waves where the identity of resonances is
less well understood.  Understandably, these methods do not have
a convenient definition for resonance fractions.  One model
introduces hadron scattering through a $K$-matrix
\cite{Wigner:km,Chung:1995dx}.
related to the $K$ and $T$-matrices and $Q$- and $F$-vectors
discussed above by
\bea
  T_{kf}(s) &=& (I-i\rho K(s))^{-1}_{ki}K_{if}(s) \nonumber \\
  Q_f(s)    &=& K^{-1}_{fk}(s)P_k(s)              \nonumber \\
  F_f(s)    &=& (I-i\rho K(s))^{-1}_{fk}P_k(s) \label{eq:kmatix}
\eea
where $\rho$ is a matrix of phase space factors, purely imaginary
below threshold, for any of the channels included. This is real,
guaranteeing the unitarity of $T$, and it contains poles and
non-resonant terms obtained from global fits to available scattering
data.  For example, Anisovich and Sarantsev \cite{Anisovich:2002ij}
derived a $K$-matrix representation of scalar $\pi\pi$ resonances
through a global fit of all the available scattering data from
threshold to 1900 MeV. The fit to the Dalitz plot then finds
parameters for the production vector, $P_k(s)$, which is, in
essence, a function that describes the dependence of production of
$f$ on $s$.  Pros and cons are known for this model.  Unlike the
isobar model, whose description of broad resonances by Breit-Wigner
is flawed, this does preserve the unitarity of the $T$-matrix.  It
does not, however, do this for $F$, since the $P$-vector is
arbitrary.  Analyticity of $T$ is not guaranteed either, so its
ability to describe poles is limited.

Less model-dependent methods are also used.  In restricted regions
of the Dalitz plot, the angular ``moments" of $f$ can be used to
measure the $S$, $P$, etc. amplitudes.  A method that works over
the whole Dalitz plot was introduced by the E791 collaboration
\cite{Aitala:2005yh}
in the analysis of $D^+\to\Km\pip\pip$ decays.  $F(s)$ for the
$\Km\pip$ $S$-wave was parameterized by a set of complex
quantities at discrete $s$ values.  These quantities were treated
as free parameters in their fit.  The method requires a reference
phase that, in their fit, was defined by the isobar model
description of the other waves.  This introduced some
model-dependence in their result.

Assumptions about the models used to describe Dalitz plots affects
results for $CKM$ phases and $\DzDzb$ mixing parameters. Extending
such model-independence into fits becomes more important, therefore,
as statistical precision of such measurements improves.

\subsubsection{Summary of Charm Dalitz plot analyses}

Several experiments have analyzed $\Dz\to\KS\pip\pim$~\cite{Yao:2006px}.
CLEO\cite{Muramatsu:2002jp} was first and included 10 resonances in
their fit:
$\KS\rho^0$, $\KS\omega$, $\KS f_{\rm 0}$(980), $\KS f_{\rm
 2}$(1270), $\KS f_{\rm 2}$(1270), $\KS h_{\rm 0}$(1370),
$K^\star{(\rm 892)}\pip$, $K^\star_0{\rm (1430)}^-\pi-$,
$K^\star_{\rm 2}{\rm(1430)}^-\pip$,  $K^\star{\rm(1680)}\pip$, and
the doubly Cabibbo suppressed mode $K^{\star}{\rm(892)}^+\pim$. CLEO
found a much smaller non-resonant contribution than did the earlier
experiments. The source of these contributions has been attributed
to broad resonances such as the $K^\star_0{\rm (1430)}$.  The
residual small non-resonant component may be a signature of broad
scalar resonances such as the $\kappares$ and $\sigmares$. More
recently, to extract information on the $CKM$ angle
$\gamma(\phi_3)$, analyses have also been made by both Belle
\cite{Poluektov:2006ia,Poluektov:2004mf} and
\babar\cite{Aubert:2005iz} using samples two orders of magnitude
larger.  At this level of statistical precision, the inadequacy of a
simple isobar model to describe the data is revealed.  After
amplitudes for a clear signal for $\rho-\omega$ mixing and for
radial excitations for the $\rho$ are added, the fit is poor.
Addition of a second, probably unphysical, $\sigma$ isobar,at $\sim
1000$~MeV/c$^2$ improves the fit yet, even then, a reasonable fit
quality is barely achieved.  Alternate isobar models lead to
uncertainties of about $10^{\circ}$ in $\gamma(\phi_3)$, but these
are smaller than the statistical and other systematic uncertainties
of about $20^{\circ}$ currently obtainable.

Decays to $\Dz\to\pip\pim\piz$ are also used by \babar\ to measure
$\gamma$ \cite{Aubert:2007ii}.  The Dalitz plot, also fitted earlier
by CLEO \cite{CroninHennessy:2005sy}, is found to have a negligibly
small $NR$ component, and to be dominated by $\rho\pi$ in all three
charge modes.  The structure of the plot shows strong, destructive
interference between these modes in a six-fold symmetry suggestive
of a dominant $I=0$.  This is consistent with the observation, noted
in section \ref{sec:had_cs}, that $\Dz\to 3\piz$ decays are strongly
suppressed.

The decay $\Dp\to\pip\pip\pim$ was studied by E687,
E791\cite{Aubert:2005sm}, FOCUS and, more recently, CLEO
\cite{Bonvicini:2007tc}.  This new CLEO analysis  uses the largest
sample ($\sim 4,000$ events) so far. An earlier isobar model
analysis by E791 had reported the need to add a $\sigma(500)$
Breit-Wigner to the $\pip\pim$ $S$-wave in order to get an
acceptable fit.  FOCUS re-examined this decay using a $K$-matrix
model for the $\pip\pim$ $S$-wave, including hadron scattering data
from a number of earlier experiments. The fit was acceptable, but
the question of whether or not there was a $\sigma(500)$ was not
clear. CLEO tried various other parameterizations for $\sigma(500)$.
Following a suggestion by J.Oller \cite{Oller:2004xm} a simple pole
of the form $1/(m_0^2-s)$ where $m_0^2=(0.47-0.22i)$ was used rather
than the Breit-Wigner used by E791. A scalar term based on the
linear sigma model \cite{Schechter:2005we} was also used. Both these
approaches produced a slightly improved fit.  This all suggests that
a low mass $\sigma$ resonance is likely to play a role in this mode,
but even larger data samples are needed to find its pole parameters.

E791 found evidence for a broad $K\pi$ scalar resonance in $\Dp\to
\Kp\pip\pip$ \cite{Aitala:1997ht}. Their original fit needed to
include a non-resonant component with a fit fraction in excess of
90\%. The inclusion of an S-wave $\bar{K}\pi$ resonance with mass
$797\pm 19\pm 43$ MeV and width $410 \pm 19 \pm 43$ MeV improved the
fit considerably and reduced the non-resonant fit fraction to $13\pm
5.8\pm 4.4$ \%.  A new fit to this channel by Focus using three times
the sample size and a $K$-matrix obtained from $K\pi$ scattering data
indicates that an $I=3/2~K\pi$ component is probably also present,
but is unable to address the existence of a $\kappa$.

The charged $S$-wave $K^{\pm}\piz$ systems could provide new
information on the $\kappa(800)$ seen, so far, only in the
neutral $\Km\pip$ system.  If it is an $I=1/2$ scalar resonance, it
could also appear, with similar mass and width, in charged $K\pi$
systems like this. This was a motivation for BaBar to study the
decays $\Dz\to\Kp\Km\piz$ \cite{:2007dc}.
An isobar model fit is able to provide a satisfactory, though ambiguous,
description of the data.  In each model tried, the $K^{*+}(890)$ resonance
is included among the components and is used as the reference for other
phases.  An $S$-wave $\Kp\Km$ resonance is required to obtain a
satisfactory fit, but there is no distinction between $a_0(980)$ and
$f_0(980)$ at low masses.  The higher mass $\Kp\Km$ system also requires
some contribution, and either $f_2'(1525)$ or an $f_0$ with a similar mass
works well.

Three models are compared for the $K\pi$
$S$-waves.  The first is a linear combination of Breit-Wigner terms for
$\kappa(800)$, $K^*(1430)$ and an $NR$ term.  The $\kappa^{\pm}$ mass
and width are allowed to vary.  Second is the model that describes the
data from the LASS $K\pi$ scattering experiment \cite{Aston:1987ir}.
The third uses the results for this wave from the E791 model-independent
$D^+\to\Km\pip\pip$ analysis \cite{Aitala:2005yh}.  The second model gives
the best fit, with the LASS phases shifted by $\sim -90^{\circ}$.
The fit with $\kappa^{\pm}$ is poor and requires a
mass $(870\pm 30)$~MeV/c$^2$ and width $(150\pm 20)$~MeV/c$^2$, the
latter differing considerably from the width of $\sim 400$~MeV/c$^2$
reported for the neutral state.   Establishing the existence of a
$\kappa$ state with such a width requires more sophisticated
analysis to find its pole in the $T$ matrix so, though poor, this
result does not rule it out as a genuine resonance.

\begin{figure}
\epsfxsize 25pc         %
\centerline{\epsfbox{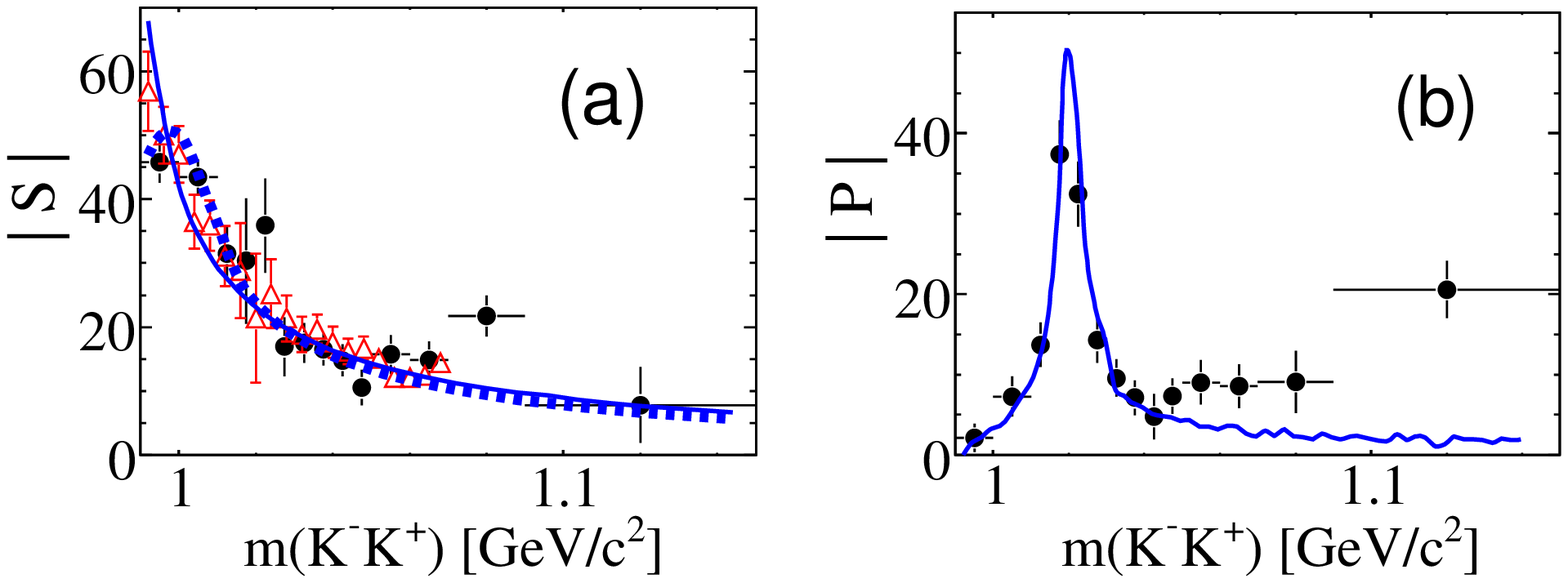}}
\caption{(a) The phase space corrected $S$-wave amplitude $|S|$ in the
 $\Kp\Km$ system, shown
 as black points, from $\Dz\to\Kp\Km\piz$ decays.  Red points are from an
 earlier analysis of $\Dz\to\Km\Kp\KS$ decays.  The solid blue curve is
 the line shape for $f_0(980)$ and the dotted curve for $a_0(980)$.  In
 (b) are black points for $|P|$ from $\Dz\to\Kp\Km\piz$ decays.  The
 solid blue curve is the $\phi$ line shape.  The figure is taken
 from Ref.~\cite{:2007dc}.
}
 \label{fig:kpkmpiz_moments}
\end{figure}

As the $\phi$ band is relatively far from other interfering vector
resonances,
a
study of the angular moments in the region around it is undertaken
in order to attempt to learn more about the underlying $S$-wave. The
$\Kp\Km$ invariant mass distributions are plotted, weighting each
event by factors $\sqrt{2\ell+\pi/4}P_{\ell}(\theta)$ (for $\ell=$0,
1 and 2), where $\theta$ is the angle between the $\Km$ and the
$\piz$ in the $\Kp\Km$ rest frame, and $P_{\ell}$ is the Legendre
polynomial function of order $\ell$. These distributions, corrected
for efficiency and with background subtracted, are defined as the
``moments" $X_{\ell}$ in each mass bin.  Assuming that only $S$- and
$P$-waves ($L=0$ and 1) contribute to the $\Kp\Km$ system, these
moments are used to extract them:
\bea
  X_0 &=& {|S|^2+|P|^2\over\sqrt 2}~;~
  X_1=\sqrt 2|S||P|\cos\theta_{SP}~;~
  X_2=\sqrt{2\over 5}|P|^2
  \label{eq:moments}
\eea
for each mass.  In Fig.~\ref{fig:kpkmpiz_moments}, the resulting
magnitudes $|S|$ and $|P|$, corrected for the phase space in the
Dalitz plot (length of the $\Kp\Km$ mass strips) are plotted.
$|P|$, shown in (b), follows the $\phi$ line shape well, with no
asymmetry and little background, up to about $1040$~MeV/c$^2$.

In Fig.~\ref{fig:kpkmpiz_moments}(a), $|S|$ values, determined from this
channel, are compared with similar measurements of $|S|$ from an earlier
analysis
\cite{Aubert:2005sm}
of $\Dz\to\Kp\Km\KS$ decays.  In that system, $a_0^+(980)$ in the
$\Kp\KS$ system was compared with the $S$-wave in the $\Kp\Km$ system
and found to agree well, suggesting that $a_0(980)$, as opposed to
$f_0(980)$ was present in both charge states.  It is possible that the
excellent agreement seen here is also evidence that $a_0(980)$ is the
main contributor to the $\Km\Kp$ $S$-wave in this decay too.  A more
convincing test would, however, be to analyze the $\Dz\to\Km\pip\eta$
and $\Dz\to\eta\piz\KS$ systems since the $a_0$ would then be more obvious
in its $\eta\pip$ decay modes.
%
%

The last topic that we discuss is the Dalitz plot analysis of the
decay $\Ds\to \Kp\Km\pip$.  E687 \cite{Frabetti:1995sg} reported
the first Dalitz plot analysis of this channel with a sample of
$\sim 300$ events.  They found significant
scalar contributions from $f_{\rm 0} {\rm(980)}$ (or $a_{\rm 0}
{\rm(980)}$).  CLEO studied the $\Kp\Km$ invariant mass spectrum in
this decay \cite{Alexander:2008cq} and found a clear peak at the
$\phi$ mass, but also a broad component with kinematic
properties indicative of a scalar.  This component is rather
important because the mode $\Ds\to \phi \pi$ has been commonly used
for $\Ds$ decay normalization, and the presence of this scalar
component introduces an additional uncertainty in this branching
fraction as the scalar channel that is absorbed in the $\phi\pi$
signal depends upon the experimental cuts used.  For example, CLEO
found that an uncertainty of the order of 5\% was introduced
depending upon the cut choices.  \babar\ \cite{Pappagallo:2007it}
recently reported a preliminary fit of the Dalitz plot structure
of this decay with over $100,000$ events.  They studied moments in
the low mass $\Kp\Km$ and $\Km\pip$ systems which showed clear
evidence for an $S$-wave contribution in the former, but none in
the latter.  An isobar model fit
shows, indeed, a strong $f_0(980)$ component, thus confirming that
the $Ds\to \phi\pi$ branching fraction is not a very wise choice of
normalization channel.

A lot of work is still ongoing on the experimental side, with
further exploration of modes relevant for \CPV\ studies, both in $D$
and $B$ decays, and on the theory side to identify tools that reduce
the model dependence in the Dalitz plot analyses, in particular when
broad scalar resonances are involved.  The rich structure of the
Dalitz plots is not only a unique and powerful asset in
understanding heavy flavor decay dynamics but also a tool for
disentangling the intricacies of non-perturbative strong interaction
effects.

\section{CHARM MIXING AND \CP-VIOLATION}\label{Mixing}

The phenomena of mixing and \CPV\ in the charm sector were first
discussed three decades ago \cite{Pais:1975qs} but the smallness of
these effects are such that experimental evidence is scarce. \CPV\
has not yet been observed, with upper limits currently at about the
1\% level. On the other hand, after years of experimental
investigation \cite{Yao:2006px}, evidence for mixing has finally
been seen in two kinds of time-dependent measurements. The \babar\
collaboration \cite{Aubert:2007wf} has reported a 3.9 standard
deviation effect in ``wrong-sign" (WS) decays of $\Dz\to\Kp\pim$
\footnote{Unless otherwise noted, charge conjugate states are
assumed throughout this paper.}.
The CDF collaboration has reported a 3.8 standard deviation effect
\cite{Aaltonen:2007uc}. Also, the Belle collaboration
\cite{Staric:2007dt} has reported a 3.2 standard deviation effect
arising from the observed difference in lifetimes for decays to \CP\
even final states $\Dz\to\Kp\Km$ and $\pip\pim$ compared to the
mixed \CP\ state $\Km\pip$.  This has also been confirmed by the
\babar\ experiment \cite{babar:dtau}. In addition, useful
information on the strong phase $\delta$ affecting $\Dz\to\Kp\pim$
mixing results is coming from CLEO-c studies that exploit the
quantum coherence of the $\Dz\Dzb$ pair produced near
threshold~\cite{Yao:2006px}.

\subsection{Charm mixing predictions in the Standard Model}

The mixing arises from $|\Delta C|=2$ interactions
that generate off-diagonal terms in the mass matrix for $D^0$ and $\Dzbar$ mesons.
The expansion of the off-diagonal terms in the neutral $D$ mass
matrix to second order in the weak interaction is
\beq
\label{M12}
\left (M - \frac{i}{2}\, \Gamma\right)_{21} =
\frac{1}{2M_D}\, \langle \Dzbar | {H}_w^{|\Delta C|=2} | D^0 \rangle
+
  \frac{1}{2M_D}\, \sum_n {\langle \Dzbar | {H}_w^{|\Delta C|=1} | n
  \rangle\, \langle n | {H}_w^{|\Delta C|=1} | D^0 \rangle
  \over M_D-E_n+i\epsilon} \ \ ,
\eeq
where ${H}_w^{|\Delta C|=2}$ and ${H}_w^{|\Delta C|=1}$
are the effective $|\Delta C|=2$ and $|\Delta C|=1$ Hamiltonians.
The off-diagonal mass-matrix terms induce mass eigenstates
$D_1$ and $D_2$ that are superpositions of the
flavor eigenstates $D^0$ and $\Dzbar$,
\begin{equation} \label{definition1}
D_{{\rm 1}\atop{\rm 2}} = p\,  D^0  \pm q\,  \Dzbar \ \ ,
\end{equation}
where $|p|^2 + |q|^2=1$.  The key quantities in $D^0$ mixing are
the mass and width differences,
\begin{equation}
\Delta M_{\rm D} \equiv M_{\rm 1} - M_{\rm 2}
\qquad  {\rm and} \qquad \Delta \Gamma_{\rm D} \equiv \Gamma_{\rm 1} -
\Gamma_{\rm 2}  \ \ ,
\label{diffs}
\end{equation}
or equivalently their dimensionless equivalents,
\beq
x_{\rm D} \equiv {\Delta M_{\rm D} \over \Gamma_{\rm D}}, \qquad {\rm and}
\qquad y_{\rm D} \equiv {\Delta \Gamma_{\rm D} \over 2\Gamma_{\rm D}}\ \ ,
\label{xy}
\eeq
where $\Gamma_{\rm D}$ is the average width of the two
neutral $D$ meson mass eigenstates.  Two quantities,
$y_{\rm D}^{\rm CP}$ and $y_{\rm D}'$,
which are actually measured in most experimental
determinations of $\Delta \Gamma_{\rm D}$, are
defined as
\bea
y_{\rm D}^{\rm CP} &\equiv& (\Gamma_+ - \Gamma_-)/
(\Gamma_+ + \Gamma_-) = y_{\rm D} \cos\phi - x_{\rm D}\sin\phi
\left(\frac{A_m}{2}-A_{prod}\right) \ \ , \nonumber \\
x' &=& x_D\cos\delta_{K\pi} + y_D\sin\delta_{K\pi} \ \ ,
\\
y' &=& y_{\rm D} \cos \delta_{K\pi} - x_{\rm D}
\sin\delta_{K\pi} \ \ ,
\nonumber
\label{y-defs}
\eea
where the transition rates $\Gamma_\pm$ pertain to decay into
final states of definite \CP,
$A_{prod} = \left(N_{D^0} - N_{{\overline D}^0}\right)/
\left(N_{D^0} + N_{{\overline D}^0}\right)$ is the so-called
production asymmetry of $D^0$ and $\overline{D}^0$ (giving
the relative weight of $D^0$ and ${\overline D}^0$ in the
sample) and $\delta_{K\pi}$ is the strong phase difference between
the Cabibbo favored and double Cabibbo suppressed
amplitudes~\cite{Falk:1999ts}.  The quantities
$A_m$ and $\phi$ account for the presence of \CPV\ in
$D^0$-${\bar D}^0$ mixing, with $A_m$ being related to the $q,p$
parameters of Eq.~(\ref{definition1}) as $A_m \equiv |q/p|^2 - 1$
and $\phi$ a \CP-violating phase of $M_{21}$ (if one neglects
direct \CPV)~\cite{Bergmann:2000id}.

The charm quark system is rather unique from the theoretical point
of view, as its mass places it somewhere on the border of heavy and
light quark systems. This makes prediction of $\DzDzb$-mixing
parameters a challenging task. As was shown in \cite{Falk:2001hx},
in the SM, $x_D$ and $y_D$ are generated only at second order in
SU(3)$_F$ breaking,
\begin{equation}
x_D\,,\, y_D \sim \sin^2\theta_C \times [SU(3) \mbox{ breaking}]^2\,,
\end{equation}
where $\theta_C$ is the Cabibbo angle.  Therefore, predicting the SM
values of $x_D$ and $y_D$ depends crucially on estimating the size
of SU(3)$_F$ breaking.

Theoretical predictions of $x_D$ and $y_D$ within the SM span
several orders of magnitude. Roughly, there are two approaches,
neither of which give very reliable results because $m_c$ is in some
sense intermediate between heavy and light.  The ``inclusive''
approach is based on the operator product expansion (OPE).  In the
$m_c \gg \Lambda$ limit, where $\Lambda$ is a scale characteristic
of the strong interactions, $\Delta M$ and $\Delta\Gamma$ can be
expanded in terms of matrix elements of local
operators~\cite{Georgi:1992as,Ohl:1992sr,Petrov:1997ch,Bigi:2000wn,Golowich:2005pt}.
Such calculations typically yield $x_D,y_D < 10^{-3}$. The use of the OPE
relies on local quark-hadron duality, and on $\Lambda/m_c$ being
small enough to allow a truncation of the series after the first few
terms.  The charm mass may not be large enough for these to be good
approximations, especially for nonleptonic $D$ decays. An
observation of $y_D$ of order $10^{-2}$ could be ascribed to a
breakdown of the OPE or of duality,  but such a large value of $y_D$
is certainly not a generic prediction of OPE analyses. The
``exclusive'' approach sums over intermediate hadronic states, which
may be modeled or fitted to experimental
data~\cite{Donoghue:1985hh,Colangelo:1990hj,Golowich:1998pz}. Since there are
cancellations between states within a given $SU(3)$ multiplet, one
needs to know the contribution of each state with high precision.
However, the $D$ meson is not light enough that its decays are
dominated by a few final states.  In absence of sufficiently precise
data on many decay rates and on strong phases, one is forced to use
some assumptions. It was shown that phase space effects alone
provide enough SU(3)$_F$ violation to induce
$x_D,y_D\sim10^{-2}$~\cite{Falk:2001hx,Falk:2004wg}. Large effects
in $y_D$ appear for decays close to $D$ threshold, where an analytic
expansion in SU(3)$_F$ violation is no longer possible; in addition,
a dispersion relation can be used to show that in this case $x_D$
would receive contributions of similar order of magnitude.

\subsection{New Physics contribution to $\DzDzb$ mixing}

In order to see how NP might affect the mixing amplitude, it is instructive to
consider off-diagonal terms in the neutral D mass matrix of Eq.~(\ref{M12}).

\subsubsection{NP in $|\Delta C|=2$ interactions.}

Since all new physics particles are much heavier than the SM ones, the most
natural place for NP to affect mixing amplitudes is in the $|\Delta
C|=2$ piece, which corresponds to a local interaction at the charm
quark mass scale. Integrating out NP degrees of freedom at some
scale $\Lambda$, we are left with an effective Hamiltonian written
in the form of series of operators of increasing
dimension\cite{Golowich:2007ka}. The complete basis of those
effective operators, which most conveniently can be done in terms of
left- and right-handed quark fields,is composed of eight operators,
\beq\label{SeriesOfOperators}
{\cal H}^{\Delta C=2}_{NP} = \sum_{i=1} C_i(\mu) ~{\cal Q}_i(\mu),
\eeq
where $C_i$ are the Wilson coefficients, and $Q_i$ are the effective operators,
\bea\label{SetOfOperators}
{\cal Q}_1 &=& \overline{u}_L \gamma_\mu c_L \overline{u}_L \gamma^\mu c_L, \qquad
{\cal Q}_5 = \overline{u}_R  \sigma_{\mu\nu} c_L \overline{u}_R \sigma^{\mu\nu} c_L
\nonumber \\
{\cal Q}_2 &=& \overline{u}_R \gamma_\mu c_R \overline{u}_L \gamma^\mu c_L, \qquad
{\cal Q}_6 = \overline{u}_R \gamma_\mu c_R \overline{u}_R \gamma^\mu c_R,
\\
{\cal Q}_3 &=& \overline{u}_L c_R \overline{u}_R c_L, \qquad\qquad
{\cal Q}_7 = \overline{u}_L c_R \overline{u}_L c_R,
\nonumber \\
{\cal Q}_4 &=& \overline{u}_R c_L \overline{u}_R c_L, \qquad\qquad
{\cal Q}_8 = \overline{u}_L  \sigma_{\mu\nu} c_R \overline{u}_L \sigma^{\mu\nu} c_R,
\nonumber
\eea
Since these operators are generated at the scale $\mu = \Lambda$ (at which NP is
integrated out), a non-trivial operator mixing can occur if we take into account
renormalization group running of these operators between $\mu=\Lambda$ and
$\mu\simeq m_c$ scales. This running can be accounted for by solving RG equations
obeyed by the Wilson coefficient functions,
\beq\label{RGE}
\frac{d}{d \log \mu} \vec C (\mu) = \hat \gamma^T(\mu) \vec C (\mu),
\eeq
where $\hat \gamma^T(\mu)$ represents the matrix of anomalous dimensions of
operators of  Eq.~(\ref{SetOfOperators})~\cite{Golowich:2007ka}.
A prediction for a mixing parameter $x$ in a particular model of NP is then
obtained by computing $C_i (\Lambda)$ for a set of ${\cal Q}_i (\Lambda)$ generated by
a given model, running the RG equations of Eq.(~\ref{RGE}) and computing matrix elements
$\langle \Dzb | {\cal Q}_i (m_c)| \Dz \rangle$.

Depending on the NP model, predictions for $x_D$ vary by orders of magnitude.
It is interesting to note that some models {\it require}
large signals in the charm system if mixing and FCNCs in the strange
and beauty systems are to be small (e.g. the SUSY alignment model).
A list of constraints on NP models is given in
Table~\ref{tab:bigtableofresults} (for more informative figures and methodology please
see Ref.~\cite{Golowich:2007ka}).
\begin{table}
\caption{Constraints on NP models from $D^0$ mixing.}
\label{tab:bigtableofresults}
\begin{center}
\begin{tabular}{ll}
\hline
Model & Approximate Constraint
\\ \hline
Fourth Generation &  $|V_{ub'} V_{cb'}|\cdot m_{b'}
<   0.5 $~(GeV)
\ \ \\
$Q=-1/3$ Singlet Quark
&  $s_2\cdot m_S  < 0.27$~(GeV)
\\
$Q=+2/3$ Singlet Quark
&  $|\lambda_{uc}| < 2.4 \cdot 10^{-4}$ \\
Little Higgs  &  Tree: See entry for $Q=-1/3$ Singlet Quark \\
& Box: Region of parameter space can reach
observed $x_{\rm D}$
\\
Generic $Z'$
&  $M_{Z'}/C > 2.2\cdot 10^3$~TeV  \\
Family Symmetries  & $m_1/f>1.2\cdot 10^{3}$~TeV
 (with $m_1/ m_2 = 0.5$)   \\
Left-Right Symmetric  & No constraint   \\
Alternate Left-Right Symmetric  &
$M_R>1.2$~TeV ($m_{D_1}=0.5$~TeV)   \\
 & ($\Delta m/m_{D_1})/M_R>0.4$~TeV$^{-1}$\\
Vector Leptoquark Bosons
& $M_{VLQ} > 55 (\lambda_{PP}/0.1) $~TeV  \\
Flavor Conserving Two-Higgs-Doublet
&   No constraint \\
Flavor Changing Neutral Higgs
&  $m_H/C>2.4\cdot 10^3$~TeV \\
FC Neutral Higgs (Cheng-Sher ansatz) &
$m_H/|\Delta_{uc}|>600$~GeV\\
Scalar Leptoquark Bosons  & See entry for RPV SUSY \\
Higgsless  & $M > 100$~TeV  \\
Universal Extra Dimensions & No constraint \\
Split Fermion
& $M / |\Delta y| > (6 \cdot 10^2~{\rm GeV})$
 \\
Warped Geometries &  $M_1 > 3.5$~TeV \\
Minimal Supersymmetric Standard  &
$|(\delta^u_{12})_{\rm LR,RL}|
< 3.5 \cdot 10^{-2}$ for ${\tilde m}\sim 1$~TeV  \\
  &
$|(\delta^u_{12})_{\rm LL,RR}| < .25 $ for ${\tilde m}\sim 1$~TeV
\\
Supersymmetric Alignment & ${\tilde m} > 2$~TeV  \\
Supersymmetry with RPV  &
$\lambda'_{12k} \lambda'_{11k}/m_{\tilde d_{R,k}} <
1.8 \cdot 10^{-3}/100$~GeV
\\
Split Supersymmetry & No constraint  \\
\hline
\end{tabular}
\end{center}
\end{table}

\subsubsection{NP in $|\Delta C|=1$ interactions.}

The local $|\Delta C|=2$ interaction cannot, however, affect $\Delta
\Gamma_{\rm D}$ because it does not have an absorptive part. Thus,
naively, NP cannot affect lifetime difference $y$. This is, however,
not quite correct. Consider a $D^0$ decay amplitude which includes a
small NP contribution, $A[D^0 \to n]=A_n^{\rm (SM)} + A_n^{\rm
(NP)}$. Here, $A_n^{\rm (NP)}$ is assumed to be smaller than the
current experimental uncertainties on those decay rates. Then it is
a good approximation to write $y$ as
\begin{eqnarray}\label{schematic}
y_D &\simeq& \sum_n \frac{\rho_n}{\Gamma_{\rm D}}
A_n^{\rm (SM)} \bar A_n^{\rm (SM)}
+ 2\sum_n \frac{\rho_n}{\Gamma_{\rm D}}
A_n^{\rm (NP)} \bar A_n^{\rm (SM)} \ \ .
\label{approx}
\end{eqnarray}
The SM contribution to $y$ is known to vanish in the limit of exact
flavor $SU(3)$. Moreover, the first order correction is also absent,
so the SM contribution arises only as a {\it second} order effect.
Thus, those NP contributions which do not vanish in the flavor
$SU(3)$ limit must determine the lifetime difference there, even if
their contributions are tiny in the individual decay
amplitudes~\cite{Golowich:2006gq}. A simple calculation reveals that
NP contribution to $y$ can be as large as several percent in
R-parity-violating SUSY models or as small as $10^{-10}$ in the
models with interactions mediated by charged Higgs
particles~\cite{Golowich:2006gq,Petrov:2007gp}.

\subsection{\CP-violation}

An observation of \CPV\ in the current round of charm experiments is arguably
one of the cleanest signals of physics beyond the SM (BSM).

It can be easily seen why manifestation of NP interactions
in the charm system is associated with the observation of (large)
\CPV. This is due to the fact that all quarks that build up the
hadronic states in weak decays of charm mesons belong to the first
two generations. Since $2\times2$ Cabibbo quark mixing matrix is
real, no \CPV\ is possible in the dominant tree-level diagrams which
describe the decay amplitudes. \CP-violating amplitudes can be
introduced in the SM by including penguin or box operators induced
by virtual $b$-quarks. However, their contributions are strongly
suppressed by the small combination of $CKM$ matrix elements
$V_{cb}V^*_{ub}$. It is thus widely believed that the observation of
(large) \CPV\ in charm decays or mixing would be an unambiguous sign
for NP. This fact makes charm decays a valuable tool in
searching for NP, since the statistics available in charm
physics experiment is usually quite large.

As with other flavor physics, \CP-violating contributions in charm can be generally
classified by three different categories:
\begin{enumerate}
\item[(I)]
\CP\ violation in the $\Delta C =1$ decay amplitudes. This type of \CPV\
occurs when the absolute value of the decay amplitude for $D$ to decay to a
final state $f$ ($A_f$) is different from the one of corresponding
\CP-conjugated amplitude (``direct \CPV''). This can happen if
the decay amplitude can be broken into at least two parts associated with
different weak and strong phases,
\beq\label{DirectAmpl}
A_f =
\left|A_1\right| e^{i \delta_1} e^{i \phi_1} +
\left|A_2\right| e^{i \delta_2} e^{i \phi_2},
\eeq
where $\phi_i$ represent weak phases ($\phi_i \to -\phi_i$ under
\CP-transformation), and $\delta_i$ represents strong phases
($\delta_i \to \delta_i$ under \CP-transformation). This ensures
that \CP-conjugated amplitude, $\overline A_{\overline f}$ would
differ from $A_f$.

\item[(II)]
\CPV\ in $\DzDzb$ mixing matrix. Introduction of
$\Delta C = 2$ transitions, either via SM or NP one-loop or tree-level NP
amplitudes leads to non-diagonal entries in the $D^0-\barD$ mass matrix,
\beq\label{MixingMatrix}
\left[M - i \frac{\Gamma}{2} \right]_{ij} =
\left(
\begin{array}{cc}
A & p^2 \\
q^2 & A
\end{array}
\right)
\eeq
This type of \CPV\ is manifest when
$R_m^2=\left|p/q\right|^2=(2 M_{12}-i \Gamma_{12})/(2 M_{12}^*-i
\Gamma_{12}^*) \neq 1$.

\item[(III)] \CPV\ in the interference of decays with and without mixing.
This type of \CPV\ is possible for a subset of final states to which
both $D^0$ and $\Dzb$ can decay.
\end{enumerate}
For a given final state $f$, \CP-violating contributions can be summarized
in the parameter
\begin{equation}
\lambda_f = \frac{q}{p} \frac{{\overline A}_f}{A_f}=
R_m e^{i(\phi_f+\delta)}\left| \frac{{\overline A}_f}{A_f}\right|,
\label{eq:lambda}
\end{equation}
where $A_f$ and ${\overline A}_f$ are the amplitudes for $D^0 \to f$ and
$\Dzb \to f$ transitions respectively and $\delta$ is the strong phase
difference between $A_f$ and ${\overline A}_f$. Here $\phi$ represents the
convention-independent weak phase difference between the ratio of
decay amplitudes and the mixing matrix.

Most of the experimental techniques that are sensitive to \CPV\ make use of
decay asymmetries, which are similar  to the ones employed in
B-physics~\cite{BigiSandaBook,Petrov:2007ms,PakvasaFirst},
\begin{eqnarray}\label{Acp}
a_f=\frac{\Gamma(D \to f)-\Gamma({\overline D} \to {\overline f})}{
\Gamma(D \to f)+\Gamma({\overline D} \to {\overline f})}.
\end{eqnarray}
One can also introduce a related asymmetry,
\begin{eqnarray}\label{Acp1}
a_{\overline f}=\frac{\Gamma(D \to \overline f)-\Gamma({\overline D} \to f)}{
\Gamma(D \to \overline f)+\Gamma({\overline D} \to f)}.
\end{eqnarray}
For charged $D$-decays the only contribution to the asymmetry
of Eq.~(\ref{Acp}) comes from the multi-component structure of the
$\Delta C =1$ decay amplitude of Eq.~(\ref{DirectAmpl}). In this case,
\begin{eqnarray}\label{DCPaF}
a_f &=& \frac{2 Im\left(A_1 A_2^*\right) \sin\delta}
{\left|A_1\right|^2 + \left|A_2\right|^2 + 2 Re A_1 A_2^* \cos\delta}
\nonumber \\
&=& 2 r_f \sin\phi_f \sin\delta,
\end{eqnarray}
where $\delta = \delta_1-\delta_2$ is the \CP-conserving phase difference and
$\phi$ is the \CP-violating one. $r_f=|A_2/A_1|$ is the ratio of amplitudes.
Both $r_f$ and $\delta$ are extremely difficult to compute reliably in
$D$-decays. However, the task can be significantly simplified if one
only concentrates on detection of NP in \CP-violating asymmetries in the
current round of experiments~\cite{Grossman:2006jg}, i.e. at the
${\cal O}(1\%)$ level. This is the level at which $a_f$ is currently probed
experimentally.
As follows from Eq.~(\ref{DCPaF}), in this case one should expect $r_f \sim 0.01$.

It is easy to see that the SM asymmetries are safely below this
estimate. First, Cabibbo-favored ($A_f \sim \lambda^0$) and doubly
Cabibbo-suppressed ($A_f \sim \lambda^2$) decay modes proceed via
amplitudes that share the same weak phase, so no \CP-asymmetry is
generated\footnote{Technically, there is a small,${\cal
O}(\lambda^4)$ phase difference between the dominant tree $T$
amplitude and exchange $E$ amplitudes.}. Moreover,
presence of NP amplitudes does not significantly change this
conclusion~\cite{Bergmann:1999pm}. On the other hand,
singly-Cabibbo-suppressed decays ($A_f \sim \lambda^1$) readily have
two-component structure, receiving contributions from both tree and
penguin amplitudes. In this case the same conclusion follows from
the consideration of the charm $CKM$ unitarity,
\beq
V_{ud} V_{cd}^* + V_{us} V_{cs}^* + V_{ub} V_{cb}^* = 0.
\eeq
In the Wolfenstein parametrization of $CKM$, the first two terms in
this equation are of the order ${\cal O}(\lambda)$ (where $\lambda
\simeq 0.22$), while the last one is ${\cal O}(\lambda^5)$. Thus,
\CP-violating asymmetry is expected to be at most $a_f \sim 10^{-3}$
in the SM. Model-dependent estimates of this asymmetry exist and are
consistent with this estimate~\cite{Buccella:1992sg}. Other
observables are also possible, such as untagged
CP-asymmetry~\cite{Petrov:2004gs}, which for some final states can
be written in terms of experimentally-measured quantities only, i.e.
which suffers no theoretical uncertainties.

\subsection{Experimental Considerations}

A number of experimental realities prevented positive evidence for
$\Dz\Dzb$ mixing until last year when sample sizes, with
sufficiently small backgrounds finally became available
\cite{Aubert:2007wf,Staric:2007dt}. Among the most important was
that the mixing rate was very small ($\RM=(\xd^2+\yd^2)/2\sim
10^{-3}$) and mixing effects were greatest at about two lifetimes
where sample sizes were quite depleted. Backgrounds were high and
use of the two way constraint on both invariant mass $M$ of the
decay products of the $\Dz$ and the difference, $\Delta M$ between
$M$ and the $D^*$ invariant mass used to tag the $\Dz$ flavor was
necessary to suppress these.  Cuts in data samples aimed at removing
$\Dz$'s from $B$ decays that adversely affect decay time
distributions also ate into sample sizes.  Also, the $\Dz$ lifetime
($\sim 400$~fs) was small on the scale of measured decay times $t$
and their uncertainties $\delta t$ and large samples were required
to properly understand their resolutions and the time offsets
arising from detector alignment issues.


\subsection{Experimental Studies of $\DzDzb$ Mixing}

Studies of  $\DzDzb$ mixing mostly use samples of $\Dz$ mesons whose
flavor is identified by the sign of the ``slow pion" $\pi_s$ in the
decay of a $D^{*+}\to\Dz\pip_s$. Effects of mixing in the subsequent
decays are then examined in one of two alternative ways. Studies of
time-dependent decay rates look either for structure in the ``wrong
sign" (WS) hadronic final states or for differences between the
lifetimes of decays to \CP\ eigenstates and to states of mixed CP.
Alternatively, time-integrated rates are studied looking for decays
either to semi-leptonic final states with the WS lepton,
\footnote{Actually, the signal for WS decays also do use information
from the time-dependence from mixed decays to distinguish them the
backgrounds.} or for the effects upon rates resulting from quantum
correlations between $\DzDzb$ systems produced in $\psi(2S)$ decays
at CLEOc.

\subsection{Time-Dependent Studies}

Mesons produced
at time $t=0$ as $\Dz(\Dzbar)$ have amplitudes for decay to final
state $f(\bar f)$ at time $t$ given by:
\bea
  \langle f|H|\Dz(t)\rangle    &=& e^{-(\gamd+i\emd)t}
         \ampl_f\left[
         \cosh((\yd+i\xd){\gamd t\over 2}) +
         \lambda_f\sinh((\yd+i\xd){\gamd t\over 2})
         \right]
  \nonumber\\
  \langle \bar f|H|\Dzbar(t)\rangle &=& e^{-(\gamd+i\emd)t}
         \bar\ampl_{\bar f}\left[
         \lambda_{\bar f}^{-1}
         \sinh((\yd+i\xd){\gamd t\over 2}) +
         \cosh((\yd+i\xd){\gamd t\over 2})
         \right]
  \nonumber\\
  \label{eq:timedep}
\eea
where $\lambda_f={q\bar\ampl_f\over p\ampl_f}$ is defined in
Eq.~(\ref{eq:lambda}), $\emd$ and $\gamd$ are the mean mass and width of
the $D_1$ and $D_2$ and $\ampl_f$ and $\bar\ampl_f$ are amplitudes
describing, respectively, the direct decays of $\Dz$ and $\Dzbar$ to
final state $f$.

In the absence of \CPV\ in either mixing or direct decay to $f$,
then $\lambda_f=e^{i\delta}$ where $\delta$ is the strong phase
difference between $\Dz\to f$ and $\Dzbar\to f$ amplitudes. With
\CPV\ in mixing ($p\ne q$) or in direct decay
($|\bar\ampl_f/\ampl_f|\ne 1$), $\lambda_f$ will be complex with
weak phase $\phi_f$ and a strong phase $\delta$.

Decay rates, proportional to the square modulus of these amplitudes
provide information on $\xd$ and $\yd$, only if the strong phase
$\delta$ is known.  This is, however, zero in the important special
case when $f$ is a \CP\ eigenstate.  In that case, if $f$ has
$CP=\eta$ the direct decay amplitudes are related by $\ampl_{\eta} =
\eta\bar\ampl_{\eta}$, so that $\delta=0$.

\subsubsection{Results from $\Dz\to\Kp\pim$ Decays.}
\label{sec:kpiws}

These WS decays can take place in two ways - either directly, by a
doubly Cabibbo suppressed (DCS) mechanism or by mixing
($\Dz\to\Dzbar$) followed by ``right sign'' (RS) Cabibbo favored
(CF) decay $\Dzbar\to\Kp\pim$. These two processes interfere giving
the time-dependent decay amplitudes given in
Eqs.~(\ref{eq:timedep}).  Neglecting \CPV\ and assuming that $|\xd|$
and $|\yd|\ll 1$), this leads to the decay rate $R_{WS}$ for these
decays \bea
  R_{WS}(t)\over e^{-\gamd t} &\propto&
            \RD+\sqrt{\RD}y't+\RM(\gamd t)^2/2
  \label{eq:mixkpi}
\eea
$\RD$ is the DCS decay rate alone, in the
absence of mixing, $\RM$ is the mixing rate and the middle term,
linear in $y'$, results from interference between mixing and DCS
amplitudes.  Deviation from a purely exponential decay, expressed in
the right side of Eq.~(\ref{eq:mixkpi}), can provide information on
$x'^2$ and $y'$, but not the values of $\xd$ and $\yd$ or their
relative signs.

This method has been used a number of times, in E691 and E791 (fixed
target experiments at Fermilab), by \babar\ and by CLEO using 9
fb$^{-1}$ of $\epm$ collisions at $\sqrt{s}\approx 10$ GeV
\cite{Godang:1999yd}.  The most stringent limit on $\Dz\Dzb$ mixing
using this method is reported by Belle \cite{Zhang:2006dp}.
Recently, however, using a 384 \ifb\ data sample, the \babar\
collaboration finally reported evidence for mixing from their sample
of $4,030 \pm 90$ decays of $\Dz$ mesons to the WS final state.

The distribution in decay times $t$ (uncertainty $\delta t$) for
these $\Dz$ and $\Dzbar$ events is shown in
Fig.~\ref{fig:babar_kpit}(a).  Also shown are the projections onto
the time axis of fits made in the $(M,\Delta M,t,\delta t)$
distribution for events near the $\Dz$ signal region.  The dashed
curve shows a fit made on the assumption that $x^{\prime
2}=y^{\prime}=0$ and the solid curve represents a fit where these
parameters are allowed to float.  Precise knowledge of parameters
describing the distribution of $\delta t$, clearly an important
component in these fits, is obtained from a sample of $\sim
1.3\times 10^6$ ``right sign" (RS) candidates for the Cabibbo
favored decays $\Dz\to\Km\pip$.  The difference between the fit with
mixing and that without is significant as seen from the residuals in
Fig.~\ref{fig:babar_kpit}(b) and in the ratio of WS to RS decays in
Fig.~\ref{fig:babar_kpit}(c).
\begin{figure}[hbtp]
  \epsfxsize 15pc
  \epsfbox{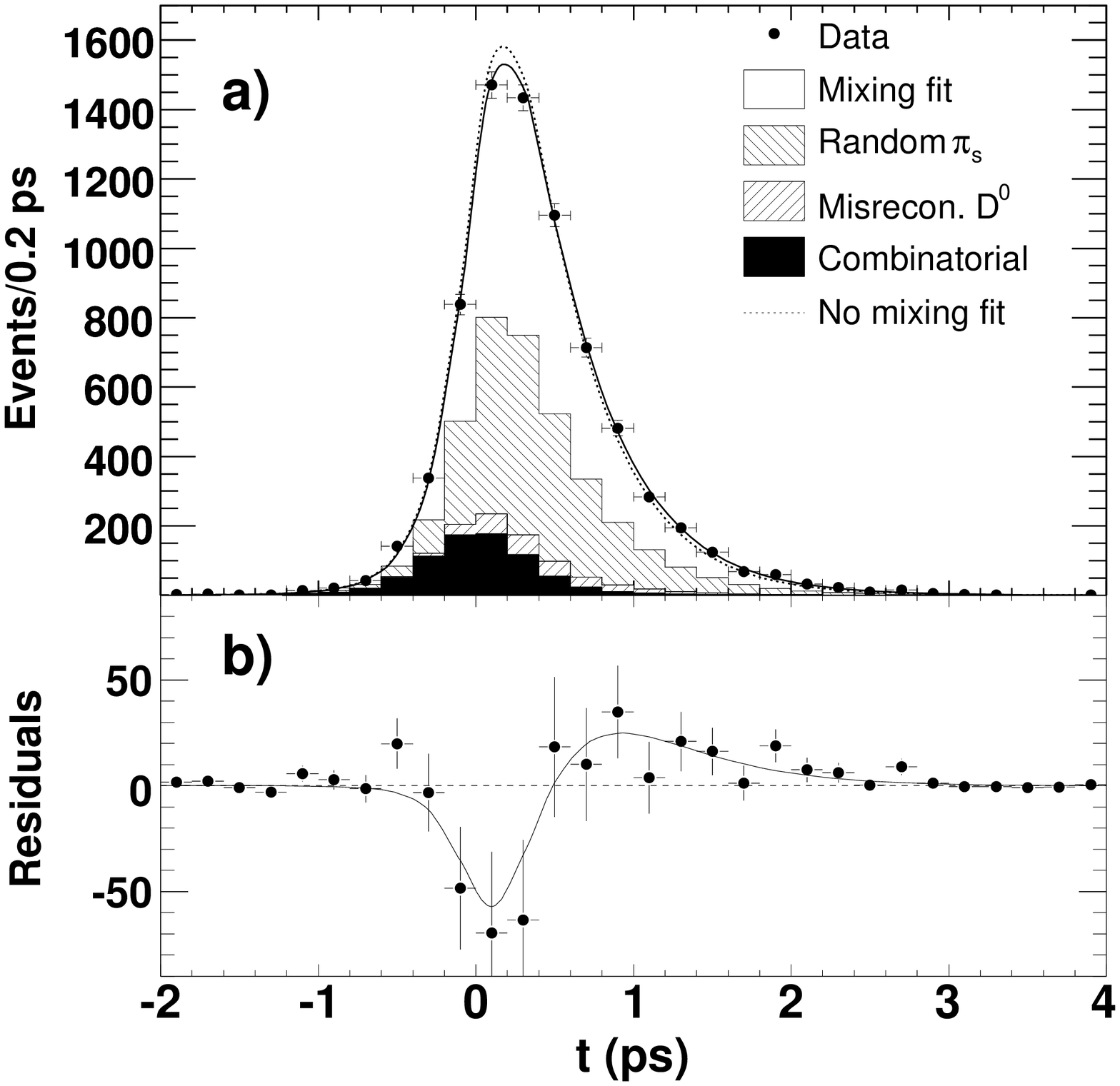}
  \epsfxsize 15pc
 \epsfbox{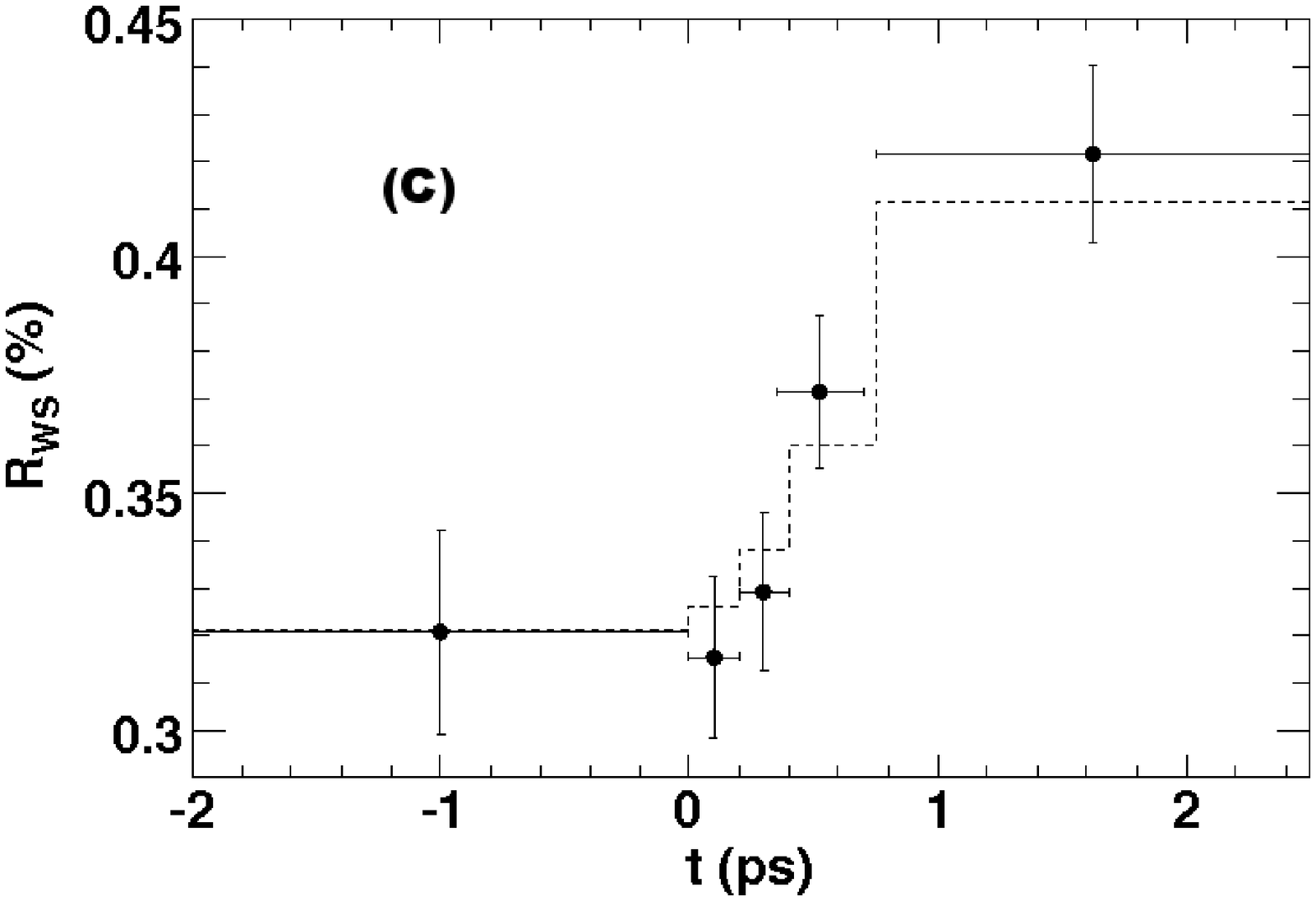}
\caption{(a) Decay times for $\Dz\to\Kp\pim$ decays.  The range
 includes negative values arising from measurement uncertainties.
 The solid curve is the result of the fit described in the text
 allowing for mixing and the dashed curve assumes no mixing
 ($x^{\prime 2}=y^{\prime}=0$).
 Contributions from various types of background are indicated.
 (b) Residuals of data bins from the fit with no mixing.  The
 solid curve is the fit with mixing.
 (c) Ratio of WS to RS decays in time slices.  The approximately
 linear form expected from Eq.~\ref{eq:mixkpi} is evident in the
 data.  Figures are from Ref.~\cite{Aubert:2007wf} (BaBar collaboration).
} \label{fig:babar_kpit}
\end{figure}
The central values from the mixing fit occur at a negative
(unphysical) value for $x'^2$.  The likelihood obtained in the fit
at the point
($x'^2=0$, $y'=6.4\times 10^{-3}$) just inside the physical region
differs from that at the central point by only 0.7 units, while the
likelihood at the no mixing point ($x'^2=y'=0$) differs by almost 24
units.  This difference is taken to indicate evidence for mixing at
$3.9`\sigma$' significance.
%
\begin{table}[hbtp]
\caption{Mixing and \CPV\ parameters from $\Dz\to\Kp\pim$ decays.}\label{tab:kpimix}
\begin{center}
\begin{tabular}{@{}lcrclrclrcl@{}}%
\toprule
Fit Type & Parameter & \mco{9}{c}{Fit Results / $10^{-3}$}    \\
         &           & \mco{3}{c}{\babar\cite{Aubert:2007wf}}
                     & \mco{3}{c}{CDF\cite{Aaltonen:2007uc}}
                     & \mco{3}{c}{Belle\cite{Zhang:2006dp}}   \\
\colrule
No \CPV\ or mixing   & $\RD$           & $3.53 $&$\pm$ & $0.09$
                                   &        &      &
                                   & $3.77 $&$\pm$ & $0.01$   \\
\hline
                 & $\RD$           & $3.03 $&$\pm$ & $0.19$
                                   & $3.04 $&$\pm$ & $0.55$
                                   & $3.64 $&$\pm$ & $0.17$   \\
No \CPV\         & $x^{\prime 2}$  &$-0.22 $&$\pm$ & $0.37$
                                   &$-0.12 $&$\pm$ & $0.35$
                                   & \mco{3}{c}{$0.18^{+0.21}_{-0.23}$} \\
                 & $y^{\prime}$    & $9.7  $&$\pm$ & $5.4 $
                                   & $8.5  $&$\pm$ & $7.6$
                                   & \mco{3}{c}{$0.6^{+4.0}_{-3.9}$}    \\
\hline
                 & $\RD$           & $3.03 $&$\pm$ & $0.19$
                                   & \mco{3}{c}{-}
                                   & \mco{3}{c}{-}            \\
                 & $\asymd$           &  $-21 $&$\pm$ & $54$
                                   & \mco{3}{c}{-}
                                   &   $23 $&$\pm$ & $47$     \\
                 & $\asymm$           & \mco{3}{c}{-}
                                   & \mco{3}{c}{-}
                                   &  $670$&$\pm$  & $1200$   \\
\CPV\ allowed    & $x^{\prime 2+}$ &$-0.24 $&$\pm$ & $0.52$
                                   & \mco{3}{c}{-}
                                   & \mco{3}{c}{$<0.72$}      \\
                 & $x^{\prime 2-}$ &$-0.20 $&$\pm$ & $0.50$
                                   & \mco{3}{c}{-}
                                   & \mco{3}{c}{-}            \\
                 & $y^{\prime+}$   & $9.8  $&$\pm$ & $7.8$
                                   & \mco{3}{c}{-}
                                   & \mco{3}{c}{$-28<y^{\prime}<21$}\\
                 & $y^{\prime-}$   & $9.6  $&$\pm$ & $7.5$
                                   & \mco{3}{c}{-}
                                   & \mco{3}{c}{-}            \\
\hline
\botrule
\end{tabular}
\end{center}
\end{table}
%
The CDF experiment, using data from a 1.5~\ifb\ $\bar pp$ exposure
at $\sqrt s=1.96$~TeV, has recently confirmed the \babar\ result at
a significance of 3.8 standard deviations~\cite{Aaltonen:2007uc}.
These results are also summarized in Table~\ref{tab:kpimix}, where
it is seen that agreement with \babar\ is excellent.  In this
experiment, it is not possible to remove background from $D$ mesons
from $B$ decays simply with a momentum cut, as in the \babar\ and
Belle experiments.  This is, however, identified from its vertex
distribution and properly taken into account in their analysis. In
an earlier analysis, the Belle collaboration \cite{Zhang:2006dp}
also obtained results with greater precision than either the \babar\
or CDF measurements.  Their central values for $x^{\prime 2}$ and
$y^{\prime}$, however, were closer to zero and so did not provide
convincing evidence for mixing and only 95\% confidence limits could
be reported.

Table~\ref{tab:kpimix} summarizes the data on mixing derived from
this hadronic channel. Generally, the highest sensitivity is
achieved with fits assuming CP violation. Note that the parameters
$x^\prime$ and $y^\prime$ include a strong phase $\delta$, the phase
of the amplitude ratio
$\langle K^+\pim|\Dz\rangle/\langle K^+\pim|\Dzb\rangle$.

\subsubsection{Lifetime Difference Measurements.}

The Belle collaboration also presented evidence for mixing using a
540~\ifb\  $\epem$ data at the $\Upsilon(4S)$.  They measured
lifetimes for singly Cabibbo-suppressed decays to the $CP=+1$ final
states $\Kp\Km$ and $\pip\pim$ and for Cabibbo-favored decays to the
final state $\Km\pip$ with mixed \CP~\cite{Staric:2007dt}.

This method was first used by the E791
collaboration~\cite{Aitala:1999dt} who were unable to detect mixing
due to limited statistical precision. Subsequently, upper limits
were reported by FOCUS, and CLEO. In the approximation that $\xd$
and $\yd$ are small, decays to $CP=+1$ eigenstates follow
approximately exponential forms with lifetimes
\cite{Bergmann:2000id}, respectively, for decays of $\Dz$ and
$\Dzbar$ of
\bea
  \label{eq:dtau}
  \tau^+ &=& \tau^0\left[1+|q/p|(y\cos\phi_f-x\sin\phi_f)\right]^{-1}
  \nonumber\\
  \tau^- &=& \tau^0\left[1+|p/q|(y\cos\phi_f+x\sin\phi_f)\right]^{-1}
\eea
where $\tau^0$ is the lifetime for decays to non-\CP\ eigenstates such as
$\Km\pip$. For such measurements, it is convenient to define
\bea
  \Delta Y = (\tau^0/\langle\tau\rangle)a_{\tau}
  \label{eq:ydy}
\eea
where $\langle\tau\rangle$ is the average of $\tau^+$ and $\tau^-$ and
$a_{\tau}=(\tau^- - \tau^+)/(\tau^- + \tau^+)$ is their
asymmetry.  In the absence of mixing ($x=y=0$) both are zero.
In the absence of \CPV\ in mixing or in decay (i.e.$\phi_f=0$)
then $\Delta Y=0$ and $\ycp=y$.

These measurements require that backgrounds are small and
have a well understood time-dependence.  The Belle samples,
consisting of $111K~\Kp\Km$, $1.22\times 10^6~\Km\pip$ and
$49K~\pip\pim$ with purities 98\% and 99\% and 92\% purity,
respectively.  The decay times for these were fit simultaneously
to distributions with an exponential for each signal convolved
with the time resolution function over the expected background
distributions.

Estimates for $\ycp$ were made using both $\Dz$ and $\Dzbar$
samples together and for $\tau^+$ and $\tau^-$ from separate
fits to each.  The major systematic uncertainties were from an
understanding of time offsets and from possible $t$-dependence of
the efficiency for reconstructing events.  The result obtained
\bea
  \ycp &=& (1.31\pm 0.32 \pm 0.25)\%
  \nonumber \\
  a_{\tau} &=& (0.01\pm 0.30 \pm 0.15)\%
  \nonumber
\eea
shows that $\ycp$ is not zero, evidence for mixing at the
$3.2\sigma$ level.  However, $a_{\tau}$ is consistent with
zero, so there is no evidence for \CPV.

A similar analysis by \babar, using a 384~\ifb\ data set, confirms
these results \cite{Staric:2007dt}. Though the \babar\ sample is
smaller, a higher purity is achieved.  Their results
\bea
  \ycp &=& (1.24\pm 0.39 \pm 0.13)\%
  \nonumber \\
  \Delta Y &=& (-0.26\pm 0.36 \pm 0.08)\%
  \nonumber
\eea
agree well with Belle's, and show evidence for mixing at the
$3.0\sigma$ level, but with no evidence for \CPV.

\subsubsection{Mixing in the Decays $\Dz\to\KS\pip\pim$.}

For $\Dz\to\KS\pip\pim$, the time dependence of the Dalitz plot
distribution allows one to measure $\xd$ and $\yd$ directly.
 This technique was first developed by the
CLEO collaboration \cite{Asner:2005sz}, who used a 9~\ifb\ data
sample.  Assuming no \CPV, they obtained the limits $(-4.7<x<8.6)$\%
and $(-6.1<y<3.5)$\% at 95 \% confidence level.

A special feature of this decay mode is that, treating $\KS$ as a
$CP=+1$ eigenstate, final states $f$ reached via the $\pi\pi$
channel are also \CP\ eigenstates for which the strong phase
difference $\delta=0$.  This provides a reference phase for a
time-dependent Dalitz plot analysis that will determine the
$\ampl_f$, $\bar\ampl_f$ for all channels everywhere in the plot.
Thus, these decays allow measurement of $\xd$, $\yd$ and their
relative sign, free from any unknown strong phase.  This fit can
also provide magnitude and phase of the $\lambda_f$'s, providing a
test for \CPV.

Belle applied this technique to a data sample 60 times larger. The
$\KS\pip\pim$ Dalitz plot   is shown in Fig.~\ref{fig:belle_kspipi_dp},
where the RS $K^{*-}$ form a vertical band and $\rho^0$ is the
diagonal band.  The WS $K^{*+}$ appears as a weak, horizontal band
destructively interfering with other structures in the plot. It is
expected that RS and WS $K^*$ amplitudes should have opposite signs
since their weak CF or DCS phases differ in sign.
\begin{figure}[hbtp]
\epsfxsize 25pc \centerline{\epsfbox{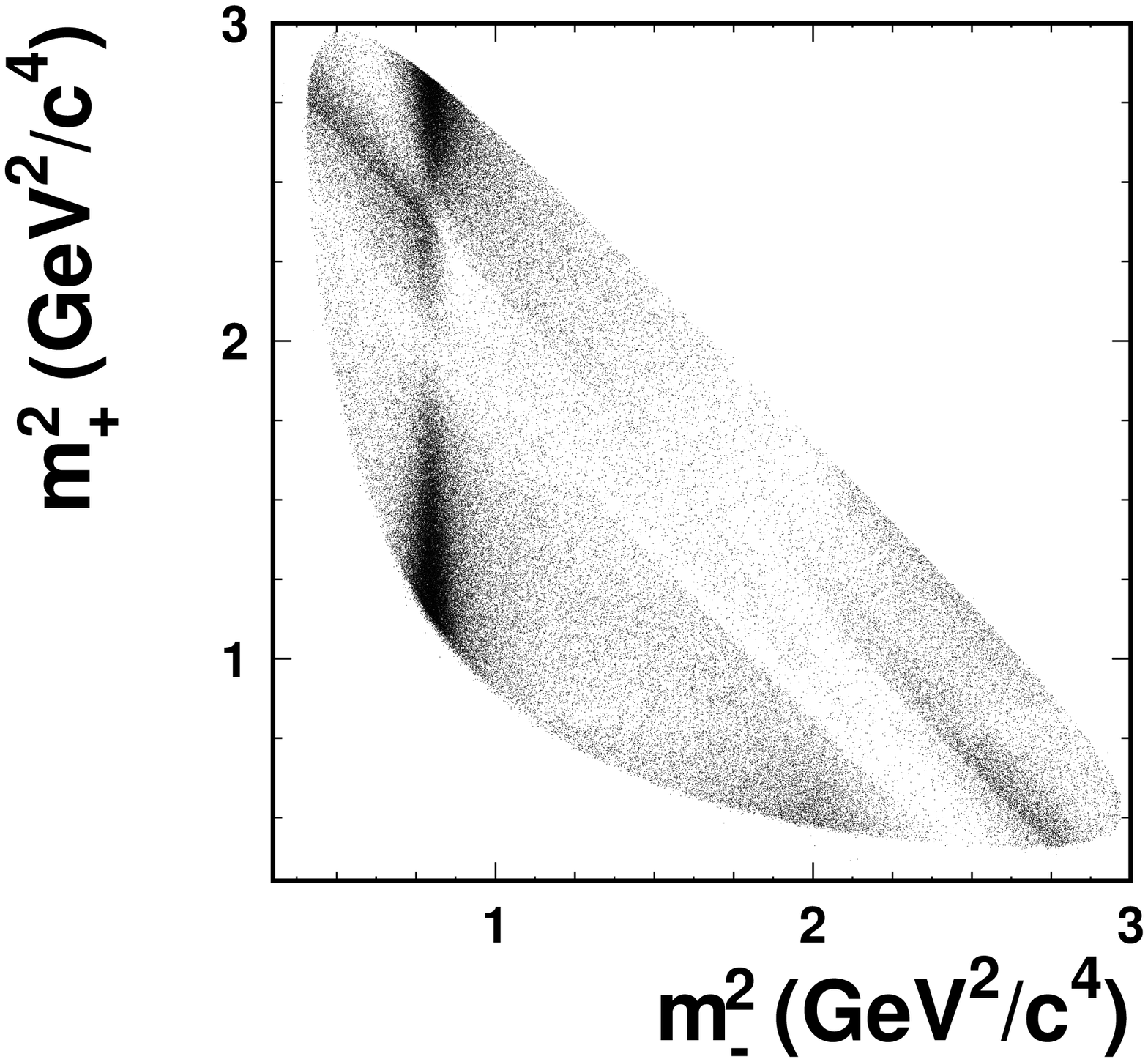}}
\caption{Dalitz plot of the decay $D \to K_S \pip\pim$
showing squared invariant mass of $\KS\pip$
 vs. that of $\KS\pim$ for data from Ref.~\cite{Abe:2007rd} (Belle collaboration).
} \label{fig:belle_kspipi_dp}
\end{figure}
An isobar model description of the amplitudes $\ampl_f$ is used,
writing them as sums of Breit-Wigner functions and spin factors,
with complex coefficients determined in the fit as in
Eq.~(\ref{eq:isobarmodel}).  For the Belle data, 18 isobars,
including $\rho-\omega$ mixing, two $\pip\pim$ $S$-wave states and a
non-resonant term $NR$ were required to provide a reasonable match
to all the features seen.  The appropriate time-dependencies were
included as in Eq.~(\ref{eq:timedep}).

Three fits were made.  In one, no \CPV\ is included (all
$\Dz$ and $\Dzbar$ events were combined and isobar coefficients
for $\Dz$ set to equal those for $\Dzbar$.  The condition $p=q$
is also imposed.  In the other fit these conditions were relaxed,
introducing a set of complex isobar coefficients for $\Dzbar$ differing
from those for $\Dz$.  The modulus and phase of the ratio $p/q$ were also
allowed to float.  In the last fit, $CPV$ in mixing ($p\ne q$) is
allowed but direct \CPV\ is not (isobar coefficients constrained to be
the same for $\Dz$ as for $\Dzbar$).  The results of these three fits
are summarized in Table~\ref{tab:kspipimix}.

%
\begin{table}[hbtp]
\caption{Mixing and \CPV\ parameters from $\Dz\to\KS\pip\pim$ decays.  The
 first uncertainty is statistical and the second systematic.  The third
 is due to uncertainties in the isobar structure assumed in the model
 for the Dalitz plot distribution.}\label{tab:kspipimix}
\begin{center}
\begin{tabular}{@{}lccc@{}}%
\toprule
Fit Type  & Parameter & Fit Result & 95\% C.L. Interval    \\
\colrule
No \CP     & $x$(\%)  & $0.80\pm 0.29\clr^{+.09}_{-.07}\clr^{+.10}_{-.14}$ &(0.0,1.6)   \\
Violation & $y$(\%)  & $0.33\pm 0.24\clr^{+.08}_{-.12}\clr^{+.06}_{-.08}$ &(-0.34,0.96)\\
\hline
          & $x$(\%)  & $0.81\pm 0.30\clr^{+.10}_{-.07}\clr^{+.09}_{-.16}$ & $|x|<1.6$  \\
\CP Viol.  & $y$(\%)  & $0.37\pm 0.25\clr^{+.07}_{-.13}\clr^{+.07}_{-.08}$ & $|y|<1.04$ \\
Allowed   & $|q/p|$  & $0.86^{+.30}_{-.29}\clr^{+.06}_{-.03}\pm 0.08 $    &            \\
          & $\arg{q/p}$ & $\left(-14^{+16}_{-18}\clr^{+5}_{-3}\clr^{+2}_{-4}\right)^{\circ}$
          \\
\hline
No direct & $|q/p|$  & $0.95^{+.22}_{-.20}$    &            \\
\CP\ viol.& $\arg{q/p}$ & $\left(-2^{+10}_{-11}\right)^{\circ}$ &            \\
\hline
\botrule
\end{tabular}
\end{center}
\end{table}
%
The authors estimate that the best solution differs from the mixing
point ($\xd=\yd=0$) by $2.2$ standard deviations.  Allowing for \CP\
violation, they obtain the \CPV\ parameters
$|q/p|=0.86^{+0.30+0.06}_{-0.29-0.03}\pm 0.08$ and
arg$(q/p=(-14^{+16+5+2}_{-18-3-4})^\circ$. This result does not
establish evidence for mixing nor for \CPV. It does, however,
illustrate a powerful way to study mixing, able to determine $\xd$,
$\yd$ and their relative signs, and the \CPV\ parameters $\phi$ and
$|p/q|$.

\subsection{Time Independent Studies.}

Rates for WS leptons in semi-leptonic decays, a clear signal for
mixing, could directly determine the mixing rate $\RM$. So far, with
samples available, these have only been able to produce upper limits
on this quantity.  The most precise limits are from Belle
\cite{Abe:2005nq}, from a 253~\ifb\  sample, and from \babar\
\cite{Aubert:2007aa} with a 344~\ifb\ sample.  The \babar\ analysis
differed considerably from Belle's in that they adopted a double
tagging approach, requiring a fully reconstructed $D$ on the side
opposite the semi-leptonic decay, used to reduce WS backgrounds.
Event yields differed by orders of magnitude between the two
experiments, yet the limits obtained were very similar:
\[\begin{array}{lc}
    \hbox{Belle}  & \RM < 1\times 10^{-3} \\
    \hbox{\babar} & (-1.3<\RM<+1.2)\times 10^{-3}
   \end{array}
\]

CLEO-c can exploit the quantum coherence of the $\Dz\Dzb$ produced
via the $\psi(3770)$ resonance to extract several important
variables affecting $\Dz\Dzb$ mixing. The pair is produced in a
$C=-1$ state, while $C=+1$ states are accessible at a higher energy,
if $\gamma\Dz\Dzb$ final states can be tagged.  By reconstructing
one neutral $D$ meson into a \CP\ eigenstate decay mode, the \CP\
eigenvalue must be the opposite for $C$ odd wave functions, and the
same for $C$ even wave functions \cite{Gronau:2001nr,Atwood:2002ak},
assuming no \CPV.  All the mixing observables can, in principle, be
measured.  A sophisticated fitting technique has been developed
\cite{Asner:2005wf} to reach maximum sensitivity.
For decays of $\psi(3770)$ to $\DzDzb$ pairs
\cite{Asner:2006md,Sun:2007fh}, the $\Dz$'s, in a
coherent $P$-wave state, and with opposite flavor and \CP, decay
in a correlated way so that $\Km\pip$ rates depend upon
$\RM$, \RD\, $y^{\prime}$ and $\delta$, the strong phase difference
between amplitudes for CF and DCS decays.  Using their 281 pb$^{-1}$
data sample collected at the $\psi(3770)$, CLEO determines
$\cos{\delta}=1.03^{+0.31}_{-0.17}\pm 0.06$ \cite{tqca}.
%

\subsubsection{Averaging the Results.}

The decays discussed above provide information in different forms,
depending upon the final state.  Semi-leptonic modes determine
$\RM$, WS hadronic systems measure $x^{\prime 2}$ and $y^{\prime}$
separately for $\Dz$ and $\Dzbar$.  Decays to \CP\ eigenstates
measure $\ycp$ and $\Delta Y$ and quantum correlated states from
$\psi(3770)\to\Dz$ decays can measure $\RM$, $\RD$, $\yd$ and
$\cos\phi$ for various hadron systems.  Parameters $\xd$, $\yd$,
$|p/q|$ and $\arg{p/q}$ are obtained from time dependence amplitude
analyses of decays of $\Dz$ to final states with more than two
hadrons, as long as those amplitudes include at least one \CP\
eigenstate.

The parameters of physics interest that define values for all these
quantities include $\xd$, $\yd$, $|p/q|$, $\arg{p/q}$,
$\phi_{K\pi}^{WS}$, $\phi_{K\pi}^{WS}$, $\RD$ and its asymmetry
$\asymd=(\RD-\bar \RD)/(\RD+\bar \RD)$.  The Heavy Flavor Averaging
Group (HFAG) made a $\chi^2$ fit to obtain values for these
parameters that best describe all 26 available observations.
\cite{schwartz:hfagmix} Results, projected onto the $(\xd,\yd)$ and
$(|p/q|,\arg{p/q})$ planes in Fig.~\ref{fig:hfag_contours}.
\begin{figure}[hbtp]
\centering
  \epsfxsize 15pc         %
  \epsfbox{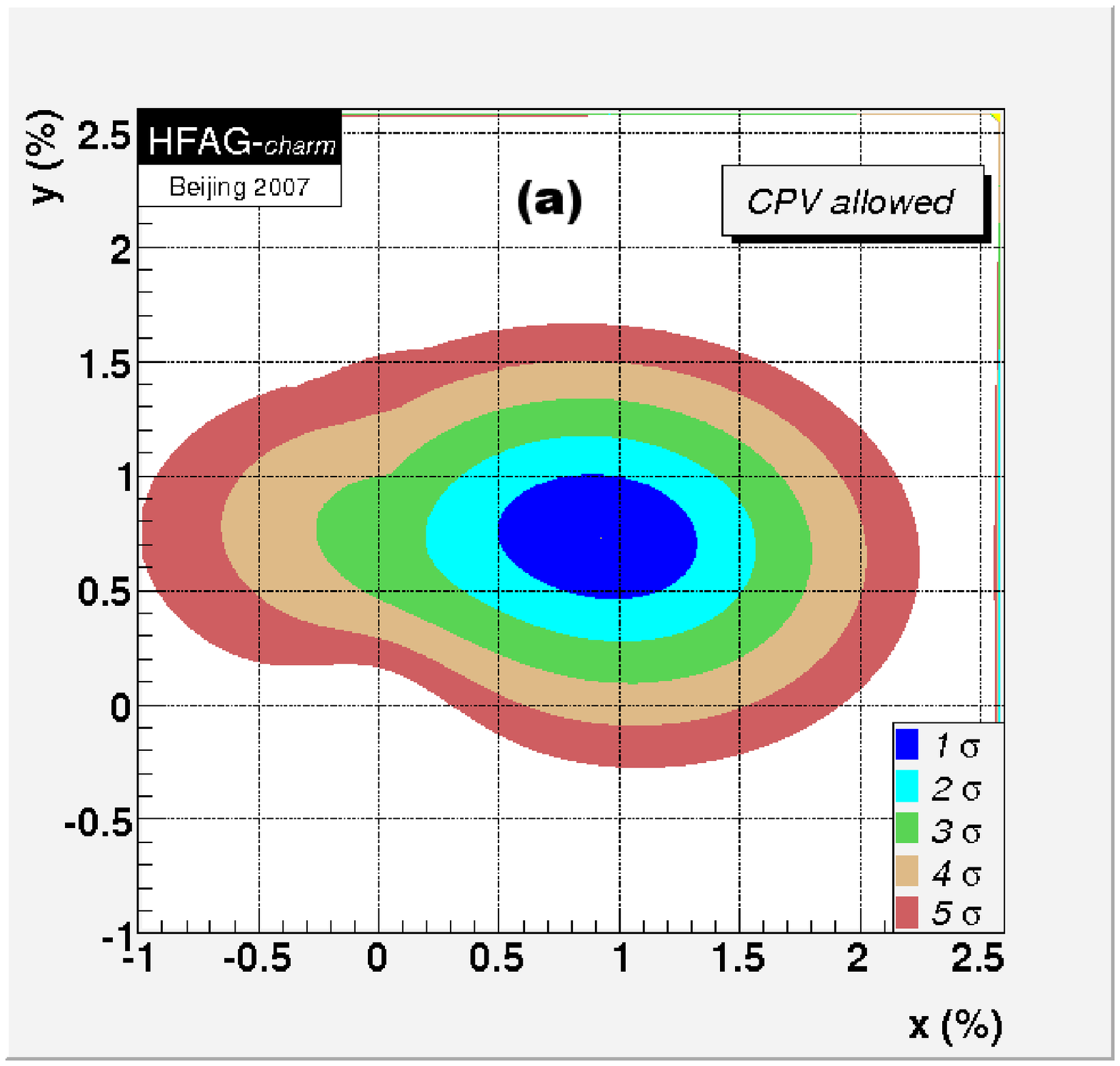}
  \epsfxsize 15pc         %
  \epsfbox{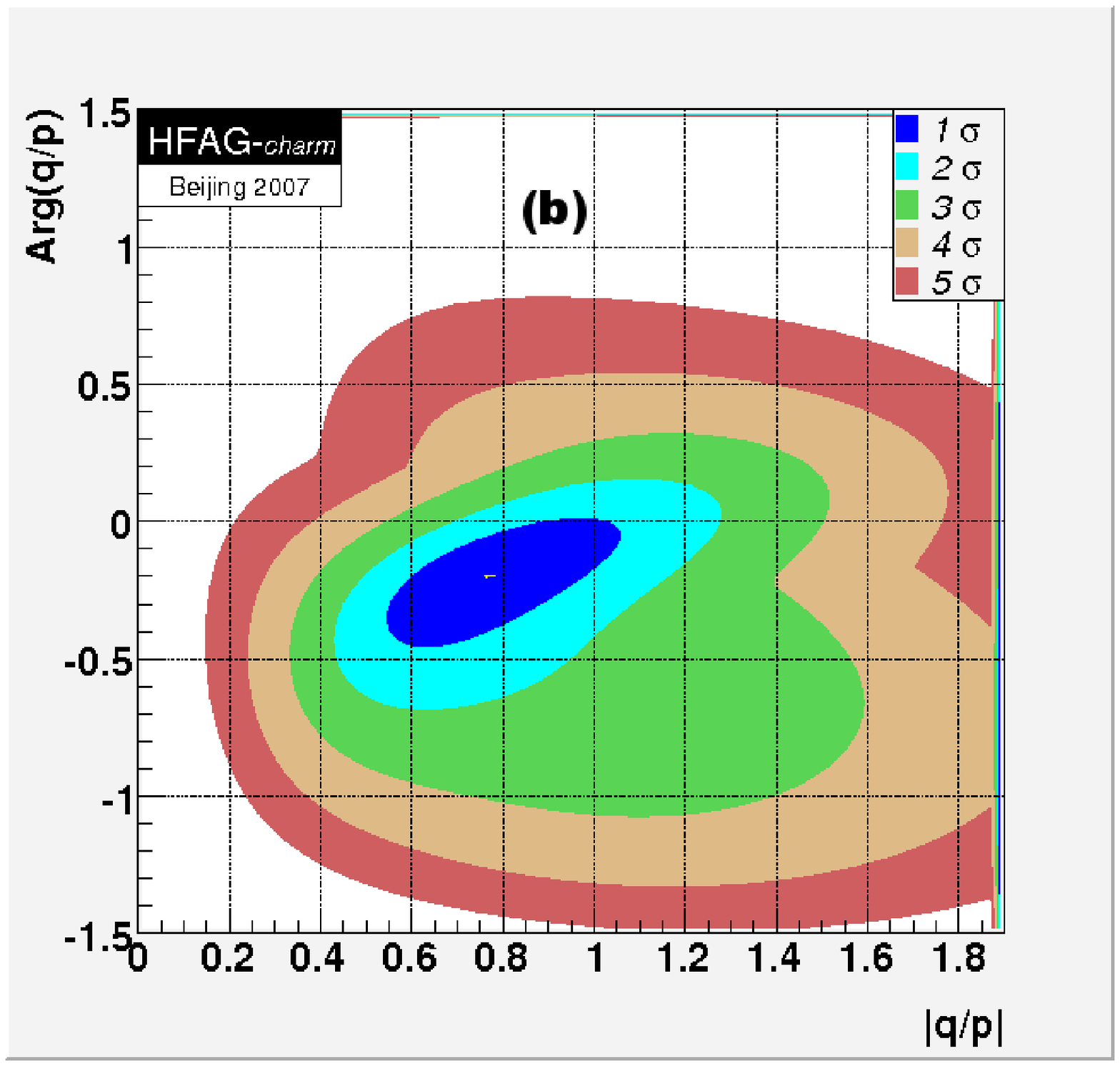}
\caption{(a) Likelihood contours projected onto $(\xd,\yd)$ plane from
 8 parameter fit to 26 mixing observables for which data exists.
 (b) Projection onto the $(|p/q|,\arg{p/q})$ plane for this fit.}
 \label{fig:hfag_contours}
\end{figure}

The point at which there is no mixing on the $(\xd,\yd)$ plane lies
at the origin, outside the $5\sigma$ contour.  This indicates very
strong evidence for $\Dz-\Dzbar$ mixing.  There is no evidence,
however, for \CPV\ in the mixing.  The point where $p=q$ in the
$(|p/q|,\arg{p/q})$ plane lies at $(1,0)$, right on the $1\sigma$
contour.

Evidence for mixing is convincing and different experiments and methods
agree well.  Since $\ycp>0$, then the $CP=-1$ state has longer lifetime,
as in the other neutral mesons that mix.  If the sign of $\xd/\yd$ remains
positive as more measurements are made, then the $CP=-1$ state is lighter,
unlike the $K^0$ system.  Finally, there is no evidence for \CPV\ so far.

\subsubsection{Results of Experimental Searches for \CPV .}\label{sec:cpv_exp}

At the present time, there is no experimental evidence for
\CPV\ in weak decays in the charm sector.  Finding it in CF or
DCS decay modes would signal possible NP.
For decays to multi-body systems, the decay asymmetry could appear
only in certain sub channels and result in particle-antiparticle
differences in phase space distribution.  Searches may also be made
for $T$-violation.

\subsubsection{Asymmetries in Time-integrated Partial Widths.}

Until this year, measurements of the
asymmetries in partial widths for time-integrated $D$ decay rates
were known with precisions of a few percent.  A recent
compilation of average values~\cite{hfag:asymmetry}
for $D$ decay asymmetries for a number of modes measured by
CDF, FOCUS, E791, CLEO, \babar\ and Belle
are summarized in Table~\ref{tab:cpv_search}.
They are all consistent with zero, with a precision of
a few \% in most cases.  Included in the most recent results
are measurements from CLEOc~\cite{:2007zt}
of CF modes using their $\psi(3770)$ data.
Precisions were excellent, of order 1\%.


%
\begin{table}[hbtp]
\caption{Asymmetries in decays of $D$ mesons to various final
 states $f$.  The asymmetry is defined in the text.}\label{tab:cpv_search}
\begin{center}
\begin{tabular}{@{}lcclc@{}}%
\toprule
\mco{1}{c}{$f$}  & \mco{1}{c}{$A_{CP}$~(\%)} &\phantom{...}&
\mco{1}{c}{$f$}  & \mco{1}{c}{$A_{CP}$~(\%)} \\
\mco{2}{c}{$\Dz$ decays:}            &&
\mco{2}{c}{$D^+$ decays:}              \\
\colrule
\mco{5}{l}{Cabibbo Favored}           \\
$\Km\pip$         & $-0.4\pm 1.0$    &&$\KS\pip$         & $-0.9\pm 0.9$ \\
$\KS\piz$         & $+0.1\pm 1.3$    &&$\KS\pip\piz$     & $+0.3\pm 0.9$ \\
$\Km\pip\piz$     & $+0.2\pm 0.9$    &&$\KS\pip\pim\pip$ & $+0.1\pm 1.3$ \\
$\Km\pip\pim\pip$ & $+0.7\pm 1.0$    &&$\Km\pip\pip$     & $-0.5\pm 1.0$ \\
                  &                  &&$\Km\pip\pip\piz$ & $+1.0\pm 1.3$ \\
\hline
\mco{5}{l}{Cabibbo Suppressed}          \\
$\Km\Kp$          & $+0.15\pm 0.34$  &&$\KS\Kp$          & $+7.1\pm 6.2$ \\
$\KS\KS$          & $-2.3\pm 1.9$    &&$\Kp\Km\pip$      & $+0.6\pm 0.8$ \\
$\pim\pip$        & $+0.02\pm 0.51$  &&$\Kp\Km\pip$      & $+0.6\pm 0.8$ \\
$\piz\piz$        & $+0.1\pm 4.8$    &&$\pim\pip\pip$    & $-1.7\pm 4.2$ \\
$\pim\pip\piz$    & $+0.1\pm  5 $    &&$\KS\Kp\pim\pip$  & $-4.2\pm 6.8$ \\
$\Km\Kp\pim\pip$  & $-8.2\pm 7.3$       \\
$\KS\pip\pim$     & $-0.9^{+2.6}_{-5.7}$\\
\hline
\mco{2}{l}{Doubly Cabibbo Suppressed}&&\mco{2}{l}{Cabibbo Suppressed \Ds}\\
$\Kp\pim$         & $-0.8\pm 3.1$    && $\Ds\to\Kp\eta$  & $-20\pm 18$   \\
$\Kp\pim\piz$     & $-0.1\pm 5.2$    && $\Ds\to\Kp\eta'$ & $-17\pm 37$   \\
$\Kp\pim\pip\pim$ & $-1.8\pm 4.4$    && $\Ds\to\KS\pip$  & $ 27\pm 11$   \\
                  &                  && $\Ds\to\Kp\piz$  & $  2\pm 29$   \\
\hline
\botrule
\end{tabular}
\end{center}
\end{table}
%
The most precise measurement comes from \babar ~\cite{Flacco:2006zz}
with a precision of 0.34\%,for the decays $\Dz\to\Kp\Km$
and of 0.5\% for $\Dz\to\pip\pim$.  This precision required
careful consideration of systematic effects, notably charge
and tagging asymmetries calibrated using data from the $\Km\pip$
CF mode.  Effects from forward-backward production asymmetry
arising from higher order QED were also taken into account.
Systematic uncertainties were 0.13\% for $\Kp\Km$ and 0.22\%
for $\pip\pim$, so the results were limited by statistics.
Nevertheless, finding \CPV\ closer to 0.1\% in CS modes, and
less in CF or DCS modes (the goals to observe NP) will
require even better precision.

\CPV\ effects would influence the Dalitz plot distributions in three-body decays
such as $\Dz\to\pip\pim\piz$ since asymmetries would affect
different partial waves in each of the three channels in differing
degrees.  Such effects would also be expected
to introduce phase differences between $\Dz$ and $\Dzbar$, and
these could be observed in analysis of the Dalitz plots.
\cite{CroninHennessy:2005sy}
found no such effects at the few \% level.  A model-independent
approach might be interesting to pursue - possibly by the $B$ factories
where an order of magnitude more data is available.
An isobar model analysis of this mode by CLEOc sees no effect
at the few percent level.

\subsubsection{$T$-violation Studies.}

It has been pointed out~\cite{BigiSandaBook}
that $T$-violation in charmed meson decays may also be
observed in the asymmetry of triple scalar products
($T$-odd) of the momenta of the particles emerging from
4-body decays~\cite{BigiSandaBook}.
The only such measurements to date are from the FOCUS
collaboration~\cite{Link:2005th},
who measured the quantities (defined in terms of momenta
$\vec p$ with suffices indicating each product from
4-body $\Dz$, $\Dzbar$ and $D_s^+$ decays)
\bea
  C_T &=& \vec p_{\Kp}\cdot
          \left(\vec p_{\pip}\times\vec p_{\pim}\right)
  \nonumber \\
  A_T &=& {\Gamma(C_T>0)-\Gamma(C_T<0)\over
           \Gamma(C_T>0+\Gamma(C_T<0)}
  \nonumber
\eea The conjugate quantities ($\bar C_T$ and $\bar A_T$) for
$\Dzbar$, $D^-$ and $D_s^-$ decays were also measured. The modes
studied were $\Dz\to\Kp\Km\pip\pim$, $D^+\to\Kp\KS\pip\pim$ and
$D_s^+\to\Kp\KS\pip\pim$. Any asymmetry not consistent with zero
would indicate $T$-violation in the absence of strong interactions.
The latter introduce the same asymmetries in particle and
anti-particle decay, so the quantity $A_{\hbox{T-viol}}=(1/2)(A_T-\bar A_T)$,
which would be zero if $T$ is conserved, is evaluated
\[\begin{array}{cr}
    \Dz\to\Kp\Km\pip\pim  &    0.010\pm 0.057\pm 0.037 \\
    D^+\to\Kp\KS\pip\pim  &    0.023\pm 0.062\pm 0.022 \\
    \Dz\to\Kp\Km\pip\pim  &   -0.036\pm 0.067\pm 0.023
  \end{array}\]
all consistent with zero.

\subsubsection{Summary.}

Clearly, experimental precision is not yet sufficient to
challenge the SM with respect to its predictions of \CPV\ in
the charm sector.  The outlook is, however, good since the
precision of asymmetry measurements is not yet limited by
systematics.  Also, model-independent studies of the
multi-body channels $\pip\pim\piz$ and $\Kp\Km\piz$, less
prone to systematic uncertainties, show promise as a tool for
observing effects of \CPV.

Both \babar\ and Belle still have large samples of 4-body decays
where $T$-violation tests similar to that made by FOCUS can be
repeated.  Also, more data is to come from both $B$ factories on the
$\Km\Kp$ and $\pim\pip$ channels where the precisions are beginning
to become interesting. Even larger samples from charm factories,
LHCb, or, possibly, Super B factories in Italy or at KEK might
produce a definitive answer on NP manifest in these studies.

\section{CONCLUSIONS AND OUTLOOK}\label{Conclusions}

Charm decays remain an exciting field for both theoretical and
experimental investigations. Charm quark transition amplitudes,
described in this review, represent a crucial tool to
understand strong interaction dynamics in the non-perturbative
regime. Complementary information that constrains model building
and lattice gauge calculations is coming from the rich spectroscopy
of charmed mesons and baryons, which is beyond the scope of this review.
The validation of theoretical tools that tackle non-perturbative
processes is critical to precision tests of the Yukawa sector of the
SM, in particular to unitarity checks of the
Cabibbo-Kobayashi-Maskawa matrix.

Finally, charm decays provide a unique window on NP, provided it
affects $u$-type quark dynamics. This way charm phenomenology can
have an impact on the interpretation of results from the direct
searches for new physics to be performed at the LHC. Charm quark is
the only $u$-type quark that can have flavor oscillations. Thus the
observation of $\Dz\Dzbar$ mixing is already constraining many
scenarios of physics beyond the SM. In addition, a multitude of new
physics models predict enhancements on \CP\ violating phases in $D$
decays, beyond the $10^{-3}$ level generally predicted within the
SM. A full exploration of \CPV\ in the charm sector, that hopefully
will be achieved in the next decade, is a critical ingredient to
further narrow the vast parameter space presently characterizing all
the new physics models.

\section{ACKNOWLEDGMENTS}\label{Acknoledgments}

The authors would like to thank their colleagues at their
institutions for helpful conversations during the course of writing
of this review, and especially D.~Asner, D.~Cinabro, R.~Harr,
S.~Pakvasa, A.~Schwartz, and S.~Stone for careful reading of the
manuscript and useful comments. The work of M.A. was supported in part by the
U.S.\ National Science Foundation under Grant PHY--0553004. B.M. was
supported by the U.S.\ National Science Foundation under Grant PHY--0457336.
A.A.P.~was supported in part by the U.S.\ National Science Foundation
CAREER Award PHY--0547794, and by the U.S.\ Department of Energy
under Contract DE-FG02-96ER41005.


\end{document}